\documentclass[11pt]{amsart}

\usepackage{amsfonts}
\usepackage{mathrsfs}
\usepackage{mathtools}
\usepackage{amssymb}
\usepackage{latexsym,amsthm}
\usepackage[all]{xy}
\usepackage{graphicx}
\usepackage{wrapfig}
\usepackage{array,multirow}
\usepackage{bm}
\usepackage{enumitem,caption}
\usepackage{mathabx} 

\usepackage[notref,notcite]{}

\usepackage{longtable}

\usepackage{wasysym}

\usepackage{hyperref}

\usepackage{tikz,tikz-cd}
\usetikzlibrary{matrix,arrows,shapes,snakes}

\newcommand{\nc}{\newcommand}
\newcommand{\rnc}{\renewcommand}

\setlist[itemize,1]{leftmargin=.4in}
\setlist[enumerate,1]{leftmargin=.4in,label=(\roman*)}
\setlist[description,1]{leftmargin=.4in,font=\normalfont\itshape}

\allowdisplaybreaks[1]

\nc{\qq}{\qquad}
\nc{\qu}{\quad}

\newtheorem{thrm}{Theorem}[section]
\newtheorem{prop}[thrm]{Proposition}

\newtheorem{lemma}[thrm]{Lemma}

\theoremstyle{definition}
\newtheorem{defn}[thrm]{Definition}

\theoremstyle{remark}
\newtheorem{rmk}[thrm]{Remark}

\nc{\rmkend}{\ensuremath{\diameter}}
\nc{\examend}{\ensuremath{\diameter}}
\nc{\defnend}{\ensuremath{\diameter}}

\numberwithin{equation}{section}

\nc{\eq}[1]{\begin{equation} #1 \end{equation}}
\nc{\eqrefs}[2]{{(\ref{#1}-\ref{#2})}}

\rnc\appendixname{}

\nc{\al}{\alpha}
\nc{\be}{\beta}
\nc{\eps}{\epsilon}
\nc{\veps}{\varepsilon}
\nc{\ga}{\gamma}
\nc{\Ga}{\Gamma}
\nc{\del}{\delta}
\nc{\Del}{\Delta}
\nc{\ze}{\zeta}
\nc{\ka}{\kappa}
\nc{\la}{\lambda}
\nc{\La}{\Lambda}
\nc{\vrho}{\varrho}
\nc{\si}{\sigma}
\nc{\Si}{\Sigma}
\nc{\vphi}{\varphi}
\nc{\om}{\omega}
\nc{\Om}{\Omega}

\nc{\A}{\mathbb{A}}
\nc{\C}{\mathbb{C}}
\nc{\F}{\mathbb{F}}
\nc{\K}{\mathbb{K}}
\nc{\N}{\mathbb{N}}
\nc{\Q}{\mathbb{Q}}
\nc{\R}{\mathbb{R}}
\nc{\Z}{\mathbb{Z}}

\nc{\mfgl}{\mathfrak{g}\mathfrak{l}}
\nc{\mfsl}{\mathfrak{s}\mathfrak{l}}
\nc{\mfb}{\mathfrak{b}}
\nc{\mfg}{\mathfrak{g}}
\nc{\mfh}{\mathfrak{h}}
\nc{\mfk}{\mathfrak{k}}
\nc{\mfn}{\mathfrak{n}}

\nc{\ot}{\otimes}

\rnc{\t}{\mathrm{t}}
\nc{\e}{\mathrm{e}}
\nc{\id}{\mathbb{I}}
\nc{\Id}{\mathrm{I}}

\DeclareMathOperator{\End}{End}

\DeclareMathOperator*{\Tr}{Tr}

\nc{\wb}{\overline}
\nc{\wh}{\widehat}
\nc{\wt}{\widetilde}

\nc{\mc}{\mathcal}
\nc{\mf}{\mathfrak}

\nc{\red}{\color{red}}
\nc{\blu}{\color{blue}}
\nc{\br}{\color{brown}}
\nc{\grn}{\color{green!55!black}}
\nc{\gry}{\color{gray}}

\nc{\ben}{\begin{eqnarray*}}
\nc{\een}{\end{eqnarray*}}
\nc{\bea}{\begin{eqnarray}}
\nc{\eea}{\end{eqnarray}}

\nc{\adag}{{a^\dag}}

\nc{\cA}{{\mathcal{A}}}
\nc{\cB}{{\mathcal{B}}}
\nc{\cC}{{\mathcal{C}}}
\nc{\cD}{{\mathcal{D}}}
\nc{\cF}{{\mathcal{F}}}
\nc{\cK}{{\mathcal{K}}}
\nc{\cM}{{\mathcal{M}}}
\nc{\cL}{{\mathcal{L}}}
\nc{\cO}{{\mathcal{O}}}
\nc{\cP}{{\mathcal{P}}}
\nc{\cQ}{{\mathcal{Q}}}
\nc{\cR}{{\mathcal{R}}}
\nc{\cS}{{\mathfrak{S}}}
\nc{\cT}{{\mathcal{T}}}
\nc{\cU}{{\mathcal{U}}}
\nc{\cV}{{\mathcal{V}}}

\nc{\ba}{{\bar a}}
\nc{\bb}{{\bar b}}
\nc{\bD}{{\wb D}}
\nc{\bcA}{\bar \cA}
\nc{\bcK}{{\wb\cK}}
\nc{\bcL}{{\wb\cL}}
\nc{\bM}{{\wb M}}
\nc{\bcM}{{\wb\cM}}
\nc{\bp}{{\wb p}}
\nc{\bcQ}{{\widebar{\cQ}}}
\nc{\bR}{{\wb R}}
\nc{\bcT}{{\wb\cT}}
\nc{\brho}{{\bar\vrho}}
\nc{\bpi}{\wb V}

\nc{\tc}{\wt c}
\nc{\tcK}{{\wt\cK}}
\nc{\tbcK}{{\wt\bcK}}
\nc{\tcL}{{\wt\cL}}
\nc{\tbcL}{{\wt\bcL}}
\nc{\txi}{\wt \xi}
\nc{\tK}{\wt K}

\nc{\tV}{\tilde{V}}
\nc{\tU}{\tilde{U}}

\newcommand{\sli}{\sum\limits}

\newcommand{\uq}{U_q(\widehat{\mathfrak{sl}}_2)}
\newcommand{\uqt}{\widetilde{U}_q(\widehat{\mathfrak{sl}}_2)}
\newcommand{\uqb}{U_q(\mathfrak{b}_+)}

\nc{\ve}{\varepsilon}
\nc{\tm}{\tilde{m}}
\nc{\Llim}{\mathrm{L}}
\nc{\Rlim}{\mathrm{R}}
\nc{\Klim}{\mathrm{K}}
\nc{\ra}{\rightarrow}

\usepackage{tikz,tikz-cd}
\usetikzlibrary{matrix,arrows,shapes,calc}
\def\Smatrix(#1,#2){
\begin{scope}[shift={(#1,#2)}]
\draw[blue,thick] (-2,0)--(0,0)--(0,-2);
\draw[red,thick] (0,2)--(0,0)--(2,-0);
\draw[red,thick,rounded corners=3] (-2,-0.3)--(-0.3,-0.3)--(-0.3,-2);
\draw[blue,thick,rounded corners=3] (0.3,2)--(0.3,0.3) --(2,0.3);
\end{scope}
}

\def\Tmatrix(#1,#2){
\begin{scope}[shift={(#1,#2)}]
\draw[blue,thick,rounded corners=3] (-2,0.15)--(-0.15,0.15)--(0,0);
\draw[blue,thick,rounded corners=3] (0,0) --(0.15,0.15)--(2,0.15);
\draw[red,thick,rounded corners=3] (-2,-0.15)--(-0.15,-0.15)--(0,0);
\draw[red,thick,rounded corners=3] (0,0) --(0.15,-0.15)--(2,-0.15);

\end{scope}
}

\def\Vmatrix(#1,#2){
\begin{scope}[shift={(#1,#2)}]
\draw[blue,thick] (-2,0)--(0,0)--(0,-2);
\draw[blue,thick] (0.3,2)--(0.3,0)--(2,0);
\draw[red,thick] (0,2)--(0,0)--(0.3,0);
\draw[red,thick,rounded corners=3] (0.3,0)--(0.3,-0.3)--(2,-0.3);
\draw[red,thick] (-2,-0.3)--(-0.8,-0.3);
\draw[red,thick,rounded corners=3] (-2,-0.3)--(-0.3,-0.3)--(-0.3,-2);
\end{scope}
}

\def\Umatrix(#1,#2){
\begin{scope}[shift={(#1,#2)}]
\draw[red,thick] (-2,0)--(-0.3,0)--(-0.3,-2);
\draw[red,thick] (0,2)--(0,0)--(2,0);
\draw[blue,thick] (-0.3,0)--(0,0)--(0,-2);
\draw[blue,thick,rounded corners=3] (0.3,2)--(0.3,0.3)--(2,0.3);
\draw[blue,thick,rounded corners=3] (-2,0.3)--(-0.3,0.3)--(-0.3,0);
\end{scope}
}

\def\Omatrix(#1,#2){
\begin{scope}[shift={(#1,#2)}]
\filldraw[fill=black!10,thick] (0,0) ellipse (0.2 and 1);
\end{scope}
}

\def\Oiso(#1,#2){
\begin{scope}[shift={(#1,#2)}]
\draw[red,thick] (-2,-0.8)--(0,-0.8);
\draw[blue,thick] (-2,-0.5)--(0,-0.5);
\draw[thick,dashed] (-2,0.8)--(0,0.8);
\draw[red,thick] (0,-0.8)--(2,-0.8);
\draw[blue,thick] (0,0.8)--(2,0.8);
\Omatrix(0,0);
\end{scope}
}

\def\Oisoinv(#1,#2){
\begin{scope}[shift={(#1,#2)}]
\draw[red,thick] (-2,-0.8)--(0,-0.8);
\draw[blue,thick] (-2,0.8)--(0,0.8);
\draw[thick,dashed] (0,0.8)--(2,0.8);
\draw[red,thick] (0,-0.8)--(2,-0.8);
\draw[blue,thick] (0,-0.5)--(2,-0.5);
\Omatrix(0,0);
\end{scope}
}

\def\lcorner(#1,#2){
\begin{scope}[shift={(#1,#2)}]
\draw[blue,thick,rounded corners=3] (0,0.7)--(0,0)--(0.7,0);
\draw[red,thick,rounded corners=3] (-0.3,0.7)--(-0.3,-0.3)--(0.7,-0.3);
\end{scope}
}

\def\rcorner(#1,#2){
\begin{scope}[shift={(#1,#2)}]
\draw[red,thick,rounded corners=3] (-0.7,0)--(0,0)--(0,-0.7);
\draw[blue,thick,rounded corners=3] (-0.7,0.3)--(0.3,0.3)--(0.3,-0.7);
\end{scope}
}

\def\Lrho(#1,#2,#3,#4){
\begin{scope}[shift={(#1,#2)}]
\draw[red,thick] (-2,0)--(2,0);
\draw (-2,0) node[left] {$#3\,$};
\draw (0,2) node[above] {$#4$};
\draw[green,thick] (0,2)--(0,-2);
\end{scope}
}

\def\Lrhobar(#1,#2,#3,#4){
\begin{scope}[shift={(#1,#2)}]
\draw[blue,thick] (-2,0)--(2,0);
\draw (-2,0) node[left] {$#3\,$};
\draw (0,2) node[above] {$#4$};
\draw[green,thick] (0,2)--(0,-2);
\end{scope}
}

\def\LOmega(#1,#2,#3,#4,#5){
\begin{scope}[shift={(#1,#2)}]
\draw[blue,thick] (-2,0)--(2,0);
\draw[red,thick] (-2,-0.3)--(2,-0.3);
\draw (-2,0.2) node[left] {$#4\,$};
\draw (-2,-0.5) node[left] {$#3\,$};
\draw (0,2) node[above] {$#5$};
\draw[green,thick] (0,2)--(0,-2);
\end{scope}
}

\def\LOmegainv(#1,#2,#3,#4,#5){
\begin{scope}[shift={(#1,#2)}]
\draw[thick,green] (-2,0) -- (2,0);
\draw[thick,red] (0,2) -- (0,-2);
\draw[thick,blue] (0.3,2) -- (0.3,-2);
\draw (-2,0) node[left] {$#3\,$};
\draw (-0.2,2) node[above] {$#4\,$};
\draw (0.5,2) node[above] {$#5$};
\end{scope}
}

\def\Ltau(#1,#2,#3){
\begin{scope}[shift={(#1,#2)}]
\draw[dashed,thick] (-2,0)--(2,0);
\draw (0,2) node[above] {$#3$};
\draw[green,thick] (0,2)--(0,-2);
\end{scope}
}

\begin{document}
\title[Cyclic Representations and Q-operators]{Cyclic Representations of $\uq$ and its Borel Subalgebras at Roots of Unity and Q-operators}
\author{Robert Weston}
\address{
Maxwell Institute for Mathematical Sciences, and Department of Mathematics, Heriot-Watt University, Edinburgh, EH14 4AS, UK}
\email{r.a.weston@hw.ac.uk}
\begin{abstract} 

We consider the cyclic representations $\Omega_{rs}$ of $\uq$ at $q^N=1$ that depend upon two points $r,s$ in the chiral Potts algebraic curve. We show how $\Omega_{rs}$ is related to the tensor product $\rho_r\ot \bar{\rho}_s$ of two representations of the upper Borel subalgebra of $\uq$. This result is analogous to the factorization property of the Verma module of $\uq$ at generic-$q$ in terms of two q-oscillator representation of the Borel subalgebra - a key step in the construction of the Q-operator. We construct short exact sequences of the different representations and use the results to construct Q operators that satisfy TQ relations for $q^N=1$ for both the 6-vertex and $\tau_2$ models.
\end{abstract}

\maketitle

\setcounter{tocdepth}{1} 
\section{Introduction}\label{sec:intro}
\subsection{Background} In 1990, Bazhanov and Stroganov published a remarkable paper \cite{BS90} in which they established a connection between two very different models: the six-vertex model and the chiral Potts model. This connection was put into a more algebraic framework and generalized in the papers \cite{DJMM91a,DJMM91b,DJMM91c,BKMS90}. The key observation is that when $q^N=1$ the algebra $\uq$ possesses $N$-dimensional so called {\it cyclic} representations in addition to the standard finite-dimensional representations inherited from $sl_2$. The cyclic representations $\Omega_{rs}$ are defined in Section \ref{sec:repns} of the current paper and depend upon a pair of points $(r,s)\in \cC_k\times \cC_k$, where $\cC_k$ is a complex algebraic curve. The R-matrix $\check{R}(rr';ss')$ corresponding to the $\uq$ isomorphism $\Omega_{rr'}\ot \Omega_{ss'}\rightarrow \Omega_{ss'}\ot \Omega_{rr'}$ was found to factorize into the composition of four $\uq$ isomorphisms, each of which swaps just one pair of the four indices $r,r',s,s'$ (see Proposition \ref{Rfact}). These four factors encode the Boltzmann weights of the chiral Potts model. 

The above work, and in particular the R-matrix factorization property, has been very influential in the field of quantum integrable systems. It became the inspiration for many papers on the factorization of R-matrices, L-operators and transfer matrices, for example \cite{BLZ96,BLZ97,De05,DKK06,De07,BJMST09,BLMS10}. The most significant of these subsequent results was the algebraic construction of Baxter's Q-operator and an understanding of how transfer matrices factor into products or combinations of products of Q-operators \cite{BLZ96,BLZ97}. A nice description of the algebraic construction of Q-operators and extensive references to the original literature can be found in the book \cite{JMS21}.

The key to understanding the algebraic construction and properties of the Q-operator for $\uq$ (generated by $e_i,f_i,t_i^{\pm 1}$, $i\in\{0,1\}$) with $q$ generic is to consider infinite-dimensional representations of the Borel subalgebra $\uqb$ generated by $ e_i,t_i^{\pm 1}$ \cite{BLZ96,BLZ97}. 
Adopting the notation of \cite{CVW24} , there are precisely two independent infinite-dimensional $\uqb$ representations $\rho_z$ and $\bar{\rho}_z$ that are {\it not} restrictions of representations of the full algebra $\uq$\footnote{The representations {\it can} be extended to the algebras known as the asymptotic algebra and shifted quantum affine algebra - see \cite{HJ12} and \cite{H20}. }. Two corresponding Q-operators, $\cQ(z)$ and $\bcQ(z)$ are defined as transfer matrices (i.e. twisted traces over the auxiliary space of products of L-operators) by choosing the auxiliary spaces as $\rho_z$ and $\bar{\rho}_z$ respectively. Throughout this work we restrict to $N$ odd and $N\geq 3$;  the representation theory for $N$ odd and $N$ even has significant differences as explained in and \cite{DCK92} and \cite{Korff_2003a}. In our story, there are also three other representations that may be chosen as auxiliary spaces: $\nu^\mu_z$ - an infinite-dimensional Verma module representation of $\uq$ that depends on a complex weight $\mu$; $\pi_z$ - the two dimensional evaluation representation of $\uq$; and $\phi_z$ - an infinite-dimensional representation of $\uqb$ which is {\it triangular} (meaning that $e_1$ acts non-trivially, but $e_0$ acts as 0). There are two relations between these different representations that give rise to corresponding properties of transfer matrices and Q-operators. They are:\vspace*{5mm}

\noindent (i) A $\uqb$ isomorphism \hspace*{2mm}
$ \nu^\mu_z \ot \phi_z \simeq \rho_{zq^{-\mu/2}}\ot \bar{\rho}_{zq^{\mu/2}}$,
\vspace*{3mm}

\noindent (ii) A short exact sequence \hspace*{2mm}
\begin{tikzcd}[cramped,sep=small]
0 \arrow[r] & \rho_{zq} \arrow[r,"\iota"] & \rho_{z} \ot \pi_{z} \arrow[r,"\tau"]  & \rho_{zq^{-1}} \arrow[r] & 0 
\end{tikzcd}\;

in which $\iota$ and $\tau$ are $\uqb$ intertwiners. 

\vspace*{3mm}
\noindent Let us denote the transfer matrices associated with the auxiliary spaces $\nu^\mu_z$ and $\pi_z$ by $T_\mu(z)$ and $T(z)$ respectively. 
The triangularity of $\phi_z$ means that the corresponding L-operator is a triangular matrix,  which in turn means that the transfer matrix with $\phi_z$ as the auxiliary space is diagonal (and in a given spin sector is proportional to the identity). Relations of the following form then follow directly from (i) and (ii) respectively (details and notation can be found in \cite{VW20,CVW24}):

\bea 
T_\mu(z) &\propto & \cQ(zq^{-\mu/2}) \bcQ(z q^{\mu/2}),\label{eq:introfact}\\
T(z) \cQ(z) &=& \alpha(z) \cQ(zq) +  \beta(z) \cQ(z q^{-1}).\label{eq:introfusion}
\eea
where $\alpha(z)$ and $\beta(z)$ are rational functions of $z$. A similar relation to (\ref{eq:introfusion}) is satisfied by $\bcQ(z)$. Relation (\ref{eq:introfact}) can be viewed as more fundamental because (\ref{eq:introfusion}) can be derived from it by considering integer $\mu$ (at which points $\nu^\mu_z$ becomes reducible).

This algebraic picture is relatively simple in the $\uq$ case discussed here because the two $\uqb$ representations $\rho_z$, $\bar{\rho}_z$ can be constructed as q-oscillator representations. This is not the case for general higher-rank algebras. However, analogous infinite-dimensional {\it pre-fundamental} representations of Borel subalgebras were constructed in \cite{HJ12} and used to extend the above picture to arbitrary untwisted quantum affine algebras. The triumphal success of this programme was to construct a full algebraic proof of the long-standing Frenkel-Reshetikhin conjectures about the form of Bethe Ansatz equations for all untwisted quantum affine algebras \cite{FH15}.  This algebraic approach has continued to develop, and is at the heart of much recent work on QQ systems and q-difference opers (see for example \cite{FH24,FJM24}). The approach has also been partially extended to open systems in \cite{FSz15,BTs18,VW20,Tsuboi2021,CVW24}. 

\subsection{Main Results}
The algebraic construction of Q-operators for $q^N=1$ has been studied in a number of papers in the 2000s, notably in \cite{Korff_2003a,Korff_2003b,Korff_2004a,Roan_2006,Roan2007}. One of the key motivations was to understand the degeneracies in the spectrum of the $XXZ$ Hamiltonian at roots of unity observed in \cite{DFM01}. 
While an algebraic understanding of the roots of unity analogue of (\ref{eq:introfusion} ) was obtained in this earlier body of work \cite{Korff_2003a},  
analogues of property (i) and Equation (\ref{eq:introfact}) in terms of representations of the Borel subalgebra were lacking. This paper seeks to fill this gap. 


The main results of this paper are as follows: after summarising the definition and properties of the $\uq$ cyclic representations of \cite{DJMM91b} in Section \ref{sec:repnspart1}, we go on in Section \ref{sec:repnspart2}  to introduce $\uqb$ cyclic representations $\rho_r$ an $\bar{\rho}_r$ for $q^N=1$ as well as another $N$-dimensional $\uqb$ representation $\varphi_c$. The later is the analog of the triangular representation $\phi_z$ mentioned above, but is even simpler in that is a direct sum of one-dimensional representations (both $e_0$ and $e_1$ act as 0). 
We then have Theorem \ref{prop:factor} that gives the $\uqb$ isomorphism 
\ben \Omega_{rs}\ot \varphi_{c_0} \simeq  \rho_{r}\ot \bar{\rho}_s.\een
 This is the analog of property (i) above. We then go on in Theorem \ref{prop:SESs} to give $\uqb$ short exact sequences that are analogous to (ii) above. Parallel results that give these factorization and fusion properties in terms of L-operators are then given in Propositions \ref{prop:Lproperties} and \ref{prop:Lfusion} respectively. In Section \ref{sec:TQ} we use these various representations as auxiliary spaces in the definition of Q operators and transfer matrices. 
 
 The construction of Q depends on whether the quantum space (the space on which T and Q act) is the tensor product of 2-dimensional representations or of $N$-dimensional cyclic representations. In the former case, we use $\rho$ and $\bar{\rho}$ as the auxiliary spaces to construct operators $Q_\rho$ and $Q_{\bar{\rho}}$ respectively. These operators satisfy the factorization property of Proposition \ref{prop:Tfact}. This can be viewed as an analog of property (\ref{eq:introfact}), but now for the transfer matrix $T_{\Omega}$ associated with the auxiliary space cyclic representation $\Omega$. The two Q-operators also satisfy the standard 6-vertex model TQ relations, as expressed in Proposition \ref{prop:TQV} and  and Equation (\ref{eq:TQstandard}).
 
 In the case when the quantum space is the tensor product of $N$-dimensional cyclic representations, the Q-operator is defined in terms of an auxiliary space representation $\Omega$ by Equation (\ref{eq:QB}), and the corresponding TQ relations are given by Proposition \ref{prop:TQW}. This later result demonstrates and explains the fact that the Q-operator of the $\tau_2$ model is indeed related to the half-monodromy matrix of the chiral Potts model as first observed in \cite{BS90}.
 
\subsection{Goals and Motivation}
As we have indicated, the overall goal of the paper is to clarify and systemize the algebraic picture of the $q^N=1$, cyclic,  $\uq$ case, and in particular to understand the role of representations of the Borel subalgebra and how they may be used to construct Q-operators that satisfy TQ relations.

There are two key motivations behind the work. Firstly, by systemizing the approach as indicated, we hope to open the way to developing the construction and properties of Q-operators for higher rank cases at $q^N=1$. Secondly, we intend to use these results in order to construct Q-operators for open systems at $q^N=1$, extending the generic-$q$ results of \cite{CVW24} to a wider class of boundary conditions and to higher rank. The finite-dimensionality of the $\uqb$ representations involved at $q^N=1$ will hopefully make this a more tractable problem that in generic $q$ case.
\newpage
\subsection{Summary of Notation}
\,\vspace*{5mm}

\begin{center}
\begin{tabular}{ | m{2.6cm} | m{10cm} | m{3cm} |  m{1.5cm} |}
\hline
{\bf Notation} & {\bf Description} & {\bf Diagrammatics}& {\bf Section} \\
\hline\hline
$q$ & primitive $N$'th root of unity, $N$ odd $\geq 3$&& \\
\hline
 $\uqt$ & central extension of quantum affine algebra $\uq$ && 2.1\\
 \hline
 $\uqb$ & Borel sualgebra of $\uqt$ && 2.2 \\
 \hline
 $W$, $V$ & $W=\C^N$, $V=\C^2$  &&  2.1.1\\
 \hline
 $X,Z$ & elements of $GL(W)$ satisfying Eqn. (\ref{eq:ZYrelns}) &&  2.1.1\\
 \hline
   $\cC_k$ & chiral Potts algebraic curve &&  2.1.1\\
   \hline
  $\Omega_{rs}$ & cyclic repn:  $\uqt\ra \End(W)$; $r,s\in \cC_k$ &
  \begin{tikzpicture}[scale=0.35]
\draw[blue,thick] (0,0)--(2,0);
\draw[red,thick] (0,-0.3)--(2,-0.3);
\draw (0,0.15) node[left] {$s\,$};
\draw (0,-0.5) node[left] {$r\,$};
\end{tikzpicture}& 2.1.1\\
\hline
 $\rho_{r}$ & cyclic repn:  $\uqb\ra \End(W)$;  $r\in \cC_k$ &
\begin{tikzpicture}[scale=0.35]
 \draw[red,thick] (0,0)--(2,0);
\draw (0,0) node[left] {$r\,$};
 \end{tikzpicture}& 2.2\\
 \hline
$ \bar{\rho}_r$ & cyclic repn:  $\uqb\ra \End(W)$;  $r\in \cC_k$ &
\begin{tikzpicture}[scale=0.35]
 \draw[blue,thick] (0,0)--(2,0);
\draw (0,0) node[left] {$r\,$};
 \end{tikzpicture}& 2.2\\

 \hline
 $\varphi_c$ & repn:  $\uqt\ra \End(W)$ (dir. sum of 1D repns); $c\in \C^\times$ &\begin{tikzpicture}[scale=0.35]
 \draw[green,thick] (0,0)--(2,0);
\draw (0,0) node[left] {$c\,$};
 \end{tikzpicture}& 2.2\\
\hline
$\pi_z$ & 2D evaluation repn:  $\uqt\ra \End(\C^2)$; $z\in \C^\times$ &\begin{tikzpicture}[scale=0.35
]
 \draw[dashed,thick] (0,0)--(2,0);
\draw (0,0) node[left] {$z\,$};
 \end{tikzpicture}& 2.2\\
 \hline
$T_{rs}$ & $\uqt$ int: $\Omega_{rs}\ra \Omega_{sr}$ & 
\begin{tikzpicture}[scale=0.35]
\Tmatrix(0,0);
\draw (-2,0.2) node[left] {$s$};
\draw (-2,-0.4) node[left] {$r$};
\draw (2,0.2) node[right] {$r$};
\draw (2,-0.4) node[right] {$s$};
\end{tikzpicture} & 2.1.2\\
\hline
$S_{rs}$ & $\uqt$ int: $:\Omega_{rr'}\ot \Omega_{ss'}\rightarrow \Omega_{rs}\ot  \Omega_{r's'} $
 & \begin{tikzpicture}[scale=0.35]
\Smatrix(0,0);
\draw (-2,0.2) node[left] {$r'$};
\draw (-2,-0.4) node[left] {$r\,$};
\draw (2,0.5) node[right] {$s'$};
\draw (2,-0.1) node[right] {$r'$};
\draw (-0.2,2) node[above] {$s$};
\draw (0.5,2) node[above] {$s'$};
\draw (-0.5,-2) node[below] {$r$};
\draw (0.2,-2) node[below] {$s$};
\end{tikzpicture}& 2.1.2\\
\hline
$\check{B}_{r';ss'}$ & $\uqt$ int $=(\id \ot T_{r's'}) S_{r's}:\Omega_{rr'} \ot \Omega_{ss'}\rightarrow \Omega_{rs} \ot \Omega_{s'r'}$ &\begin{tikzpicture}[scale=0.35]
\Vmatrix(0,0); 
\draw (-2,-0.4) node[left] {$r\,$};
\draw (-2,0.3) node[left] {$r'$};
\draw (-0.2,2) node[above] {$s$};
\draw (0.5,2) node[above] {$s'$};
\draw (-0.5,-2) node[below] {$r$};
\draw (0.2,-2) node[below] {$s$};
\draw (2,-0.4) node[right] {$s'$};
\draw (2,0.3) node[right] {$r'$};
\end{tikzpicture}
& 2.1.2\\
\hline
$\check{A}_{rr';s}$ & $\uqt$ int $=S_{rs}( T_{rr'}\ot \id):\Omega_{rr'}\ot \Omega_{ss'}\rightarrow. \Omega_{r's}\ot \Omega_{rs'}$ &\begin{tikzpicture}[scale=0.35]
\Umatrix(0,0); 
\draw (-2,0.6) node[left] {$r'$};
\draw (-2,-0.2) node[left] {$r\,$};
\draw (2,0.6) node[right] {$s'$};
\draw (2,-0.2) node[right] {$r$};
\draw (-0.2,2) node[above] {$s$};
\draw (0.5,2) node[above] {$s'$};
\draw (-0.4,-1.9) node[below] {$r'$};
\draw (0.2,-2.2) node[below] {$s$};
\end{tikzpicture}
& 2.1.2\\
\hline
$\check{R}(rr';ss')$&  $\uqt$ int $=S_{rs'} (T_{rs}\ot T_{r's'}) S_{r's}:\Omega_{rr'}\ot \Omega_{ss'}\ra \Omega_{ss'}\ot \Omega_{rr'}$ &see text&2.1.2\\
\hline
$\overline{W}_{rs}(n), \,\widehat{W}_{rs}(n)$ & Components of $T_{rs}$ and $S_{rs}$ = Chiral Potts Boltz. wts && 2.1.2\\
\hline
$W_{rs}(n)$, $\widecheck{W}_{rs}(n)$ & Transformed versions of $\overline{W}_{rs}(n)$, $\widehat{W}_{rs}(n)$ && 2.1.2\\
\hline
$\chi$ & $=X^{-1}\ot X$&& 3.1\\
\hline
$\cO(\chi)$ & $\uqb$ int $: \Omega_{rs}\ot \varphi_{c_0} \rightarrow \rho_{r}\ot \bar{\rho}_s$ &
\begin{tikzpicture}[scale=0.3]
\Oiso(0,0);
\draw (-2,-0.9) node[left] {$r$};
\draw (-2,-0.3) node[left] {$s$};
\draw (2,-0.9) node[right] {$r$};
\draw (2,1) node[right] {$s$};
\end{tikzpicture}
& 3.1\\
\hline
$\cP(Z)$ & polynomial $=\sum\limits_{n=0}^{N-1} p_n Z^{2n}$ && 3.1\\
\hline
$\mathfrak{T}_{rs}$ & $\uqb$ int $:\rho_r\ot \bar{\rho}_s \rightarrow \rho_s \ot \bar{\rho}_r, $ && 3.1\\
\hline
$\cS_{rs}$ & $\uqb$ int $:\bar{\rho}_r\ot \rho_s \rightarrow \bar{\rho}_s\ot \rho_r.$ && 3.1\\
\hline
\end{tabular}
\end{center}
\begin{center}
\begin{tabular}{ | m{2.6cm} | m{10cm} | m{3cm} |  m{1.5cm} |} 
\hline
{\bf Notation} & {\bf Description} & {\bf Diagrammatics}& {\bf Section} \\
\hline\hline
$\iota_s$ & $\uqb$ int $:\rho_{sq} \ra  \rho_{s} \ot \pi_{z_s}$ &
\begin{tikzpicture}[scale=0.4]
\draw[red,thick] (0,0)--(2,0);
\draw[thick,green,rounded corners=12] (1,0) -- (1,1) -- (2,1);
\draw (0,0) node[left] {$sq\,$};
\draw (2,0) node[right] {$s\,$};
\draw (2,1) node[right] {$z_s\,$};
\end{tikzpicture}
& 3.2\\
$\bar{\iota}_s$ & $\uqb$ int $:\bar{\rho}_{sq} \ra  \bar{\rho}_{s} \ot \pi_{z_s}$ &see text & 3.2\\
$\bar{I}_s$ & $\uqb$ int $:\Omega_{r,sq} \ra  \Omega_{r,s} \ot \pi_{z_s}$ &see text & 3.2\\
\hline
$\tau_s$ & $\uqb$ int $:\rho_{s} \ot \pi_{z_s} \ra \rho_{sq^{-1}}$&
\begin{tikzpicture}[scale=0.4]
\draw[red,thick] (0,0)--(2,0);
\draw[thick,green,rounded corners=12] (0,1) -- (1,1) -- (1,0);
\draw (0,0) node[left] {$s\,$};
\draw (2,0) node[right] {$sq^{-1}\,$};
\draw (0,1) node[left] {$z_s\,$};
\end{tikzpicture}
& 3.2\\
$\bar{\tau}_s$ & $\uqb$ int $:\bar{\rho}_{s} \ot \pi_{z_s} \ra \bar{\rho}_{sq^{-1}}$&see text & 3.2\\
$\bar{T}_s$ & $\uqb$ int $:\Omega_{r,s} \ot \pi_{z_s} \ra \Omega_{r,sq^{-1}}$&see text & 3.2\\
\hline
$\check{L}_{\Omega_{rs}}(z)$ & $\uqt$ int $:\Omega_{rs}\ot \pi_z\rightarrow \pi_z\ot \Omega_{rs}$ & \begin{tikzpicture}[scale=0.3]
\LOmega(0,0,r,s,z);
\end{tikzpicture} & 3.3\\
\hline
$\check{L}_{\varphi}$ & $\uqb$ int $:\varphi_c\ot \pi_z\rightarrow \pi_z\ot \varphi_c$ & \,\,\,\,\,
 \begin{tikzpicture}[scale=0.3]
\Ltau(,0,z)
\end{tikzpicture} & 3.3\\
\hline
$\check{L}_{\rho_{r}}(z)$& $\uqb$ int   $:\rho_{r}\ot \pi_z\rightarrow \pi_z\ot \rho_{r}$   & \begin{tikzpicture}[scale=0.3]
\Lrho(0,0,r,z);
\end{tikzpicture} & 3.3\\
\hline
 $\check{L}_{\bar{\rho}_{r}}(z)$& $\uqb$ int  $:\bar{\rho}_{r}\ot \pi_z\rightarrow \pi_z\ot \bar{\rho}_{r}$ & \begin{tikzpicture}[scale=0.3]
\Lrhobar(0,0,r,z);
\end{tikzpicture} & 3.3\\
\hline
$\check{\bf L}_{\Omega_{rs}}(z)$& $\uqt$ int $:\pi_z\ot \Omega_{rs} \rightarrow \Omega_{rs}\ot \pi_z$ & \begin{tikzpicture}[scale=0.3]
\LOmegainv(0,0,z,r,s);
\end{tikzpicture} & 3.3\\
\hline
$\check{R}(z/w)$ & $\uqt$ int  $:  \pi_z\ot \pi_w \rightarrow \pi_w\ot \pi_z$ ; 6-vertex model R-matrix & \begin{tikzpicture}[scale=0.3]
\draw[green,thick] (-2,0)--(2,0);
\draw[green,thick] (0,2)--(0,-2);
\draw (-2,0) node[left] {$z$};
\draw (0,2) node[above] {$w$};
\end{tikzpicture}
& 3.3\\
\hline
$U_r(z)$, $V_r(z)$ & $2\times 2$ matrices; $r\in\cC_k$, $z\in \C^\times$ && 3.3\\
\hline
$\{A,B\}$ & Element of $\End(W\ot V)$ defined by Eqn (\ref{eq:ABnotn}) & & 3.3\\
\hline
$T(z)$ & $\in \End(V^{\ot M})$ ; 6-vertex model transfer matrix &see text& 4.1\\
\hline
$Q_{\rho_r}(w)$ & $\in \End(V^{\ot M})$ ; 6-vertex model Q-operator &see text& 4.1\\
\hline
$Q(z,\mu)$ & transformed version of $Q_{\rho_r}(w)$ & see text& 4.1\\
\hline
$Q_{\bar{\rho}_r}(w)$ & $\in \End(V^{\ot M})$ ;  6-vertex model Q-operator &see text& 4.1\\
\hline
$\widebar{Q}(z,\mu)$ & transformed version of $Q_{\bar{\rho}_r}(w)$ & see text& 4.1\\
\hline
$T_{\Omega_{rs}}(w)$& $\in \End(V^{\ot M})$ &see text& 4.1\\
\hline
$T_{\varphi_{c}}(w)$& $\in \End(V^{\ot M})$ &see text& 4.1\\
\hline
$\cT(z)$ & $\in \End(W^{\ot M})$ ;  $\tau_2$ model  transfer matrix&see text& 4.2\\
\hline
${\cQ}_{r';ss'} $ &$\in \End(W^{\ot M})$ ;  $\tau_2$ model Q-operator &see text& 4.2\\
\hline
$\cT^{CP}$  & $\in \End(W^{\ot M})$ ; chiral Potts model  transfer matrix&see text& 4.2\\

\hline

\end{tabular}
\end{center}

\newpage

\section{Cyclic Representations of $\uqt$ and $\uqb$}\label{sec:repns}
In this Section, we recall the cyclic representations of $\uqt$ defined in \cite{DJMM91b} and their connection with the Boltzmann weights of the chiral Potts model. We shall then go on to define new cyclic representations of the Borel subalgebra $\uqb$, as well as a fully decomposable $\uqb$ representation $\varphi_c$. Throughout this paper we fix $q$ to be a primitive $N$'th root of unity, with $N$ odd and $N\geq 3$.

\subsection{Cyclic Representations and R-matrices of $\uqt$}\label{sec:repnspart1}
We closely follow the notation and conventions of Section 4 of \cite{DJMM91b} who in turn developed the earlier work of \cite{DCK92} on representations  of quantum groups at roots of unity.
We make use the algebra $\uqt$ defined in Section 4 of \cite{DJMM91b}: it is an algebra over $\mathbb{C}$ generated by $e_i,f_i,t_i^{\pm 1},z_i$ $(i=0,1)$, where $e_i,f_i,t_i^{\pm 1}$ satisfy the standard relations of the quantum affine algebra $\uq$, and $z_0$ and $z_1$ are two new central elements. The comultiplication of $\uqt$ is chosen as 
\ben \Delta(e_i)&=&e_i\ot \id + z_i t_i\ot e_i,\quad \Delta(f_i)=f_i\ot t_i^{-1} + z_i^{-1}\ot f_i,\\
\Delta(t_i)&=&t_i \ot t_i,\quad \Delta(z_i)=z_i\ot z_i.\een  
\subsubsection{Cyclic Representations $\Omega_{rs}$}\label{sec:Omegadef}
Let $W$ be an $N$-dimensional vector space over $\C$, and let $X,Z$ be invertible linear operators on $W$ satisfying the relations \bea ZX=qXZ,\quad X^N=Z^N=\id.\label{eq:ZYrelns}\eea
Following \cite{DJMM91b}, we define cyclic representations $\Omega_{rs}:\uqt\rightarrow \End(W)$ that depend on a pair of points $(r,s)\in \cC_k\times \cC_k$. Here, $\mathcal{C}_k$ is the complex algebraic curve given by $(x,y,\mu)\in (\C^\times)^3$ such that
\ben
x^N+y^N=k(1+x^N y^N) \,,
\quad
\mu^N=\frac{k'}{1-k x^N}=\frac{1-ky^N}{k'} \,,\;\;
\een
where $k$ is the modulus with $k^2+k'^2 = 1$. With the notation $r=(x_r,y_r,\mu_r)$ and $s=(x_s,y_s,\mu_s)$, the representation $\Omega_{rs}:\uqt\rightarrow \hbox{End}(W)$ is given by
\footnote{
To avoid confusion, we note that our $\Omega_{rs}$ corresponds to $\xi=sr$ in the notation of \cite{DJMM91b}. The other difference in notation is that $Z$ of \cite{DJMM91b} is our $Z^2$ - the reason for this change will become apparent when we consider L-operators in Section \ref{sec:Lops}.
}
\begin{equation} \label{eq:cyclrep}
  \begin{aligned}
    & \Omega_{rs} (e_0)= \frac{\kappa_0 x_r}{q-q^{-1}} X^{-1} \left(\frac{y_s}{x_r\mu_r\mu_s} Z^{-2}-1\right), \,\quad  \Omega_{rs} (e_1)= \frac{\kappa_1 y_r}{q-q^{-1}} \left(\frac{x_s\mu_r\mu_s}{y_r} Z^2-1\right)X ,\\
    & \Omega_{rs} (f_0)= \frac{c_0 y_r}{q\kappa_0 x_r x_s (q-q^{-1})} \left(\frac{x_s\mu_r\mu_s}{y_r} Z^2-1\right)X  ,\quad  \Omega_{rs} (f_1)= \frac{c_0}{q \kappa_1 x_s(q-q^{-1})} X^{-1} \left(\frac{y_s}{x_r\mu_r\mu_s} Z^{-2}-1\right)  ,\\
    &  \Omega_{rs} (t_0)= \frac{c_0 y_r y_s}{q^2 x_r x_s\mu_r\mu_s} Z^{-2},\quad  \Omega_{rs} (t_1)= (\Omega_{rs}(t_0))^{-1}
    ,\quad  \Omega_{rs} (z_0)=c_0,\quad  \Omega_{rs} (z_1)=\frac{1}{c_0},
  \end{aligned}
\end{equation}
where \bea \displaystyle c_0^2=\frac{q^2 x_r x_s}{y_r y_s},\label{eq:c0}\eea and the values of the fixed parameters $\kappa_0,\kappa_1$ can be found in \cite{DJMM91b}. When it clarifies the expression we sometimes use the notation $\Omega_{r,s}$ instead of $\Omega_{rs}$.
\subsubsection{R-Matrices $\check{R}(rr';ss')$}
The $\uqt$ representation $\Omega_{rs}$ was constructed, and parameterized in terms of $(r,s)\in \cC_k\times \cC_k$, precisely in order that an invertible $\uqt$ linear map \bea\check{R}(rr';ss'):W\ot W \rightarrow W\ot W\label{eq:WWaction}\eea existed with the property
\bea \check{R}(rr';ss') (\Omega_{rr'}\ot \Omega_{ss'}) \Delta(x) = (\Omega_{ss'}\ot \Omega_{rr'}) \Delta(x) \check{R}(rr';ss'),\quad \hbox{for all } x\in \uqt.\label{eq:Omegacomm}\eea

As we shall be discussing many such intertwiners in this paper, it is useful at this point to introduce some abbreviated notation:
the pair of properties (\ref{eq:WWaction}) and (\ref{eq:Omegacomm}) is of course the same as the statement that there exists a isomorphism of $\uqt$ representations
 \begin{eqnarray} \check{R}(rr';ss'):(\Omega_{rr'},W) \ot (\Omega_{ss'},W) \ra (\Omega_{ss'},W) \ot (\Omega_{rr'},W), \label{eq:intt}\end{eqnarray}
 where the representation $(\Omega_{rs},W)$ consists of the vector space $W$ and the algebra action $\Omega_{rs}:\uqt \ra \End(W)$. Our abbreviation is to express relation (\ref{eq:intt}) as
 \begin{eqnarray} \check{R}(rr';ss'):\Omega_{rr'} \ot \Omega_{ss'} \ra \Omega_{ss'}\ot \Omega_{rr'}.\label{eq:intttt}\end{eqnarray}


The key factorization property of the R-matrix that was discussed in Section \ref{sec:intro} is as follows.
\begin{prop}[ Date, Jimbo, Miki, Miwa]\label{Rfact} The $\uqt$ isomorphism $\check{R}(rr';ss'):\Omega_{rr'}\ot \Omega_{ss'}\ra \Omega_{ss'}\ot \Omega_{rr'}$ can be written in terms of two $\uqt$ isomorphisms \ben T_{rs}:\Omega_{rs}&\rightarrow&\Omega_{sr}, \hbox{and}\\
S_{r's}:\Omega_{rr'}\ot \Omega_{ss'} &\ra &\Omega_{rs}\ot \Omega_{r's'},\een as follows
\bea 
\check{R}(rr';ss')= S_{rs'} (T_{rs}\ot T_{r's'}) S_{r's},\label{eq:Rfact}\eea
where 
\bea T_{rs}=\sum_{n=0}^{N-1}\overline{W}_{rs}(n) Z^{2n},\quad
S_{rs}=\sum_{n=0}^{N-1}\widehat{W}_{rs}(n)
 \chi^n,\quad \chi:=X^{-1}\ot X,\label{eq:TSform}\eea
 in which 
 the coefficients $\overline{W}_{rs}(n)$ and $\widehat{W}_{rs}(n)$ are chosen to satisfy the recursion relations
\bea \frac{\overline{W}_{rs}(n)}{\overline{W}_{rs}(n-1)}= \frac{\mu_r\mu_s(x_r q^2 -x_s q^{2n})}{y_s-y_r q^{2n}},\quad 
\frac{\widehat{W}_{rs}(n)}{\widehat{W}_{rs}(n-1)}= \frac{\mu_s y_r-\mu_r y_s q^{2(n-1)}}{\mu_s x_s-\mu_r x_r  q^{2n}}.\label{eq:WRR} \eea
\end{prop}
\begin{proof}
    This statement is Theorem 4.2 of  \cite{DJMM91b} and the full proof can be found there. Here, we note that 
   assuming $T_{rs}$ and $S_{rs}$ are of the form (\ref{eq:TSform})
 then it follows that they are $\uqt$ intertwiners of the form stated provided they satisfy respectively
      \bea
       T_{rs} (y_r-x_s \mu_r \mu_s Z^2) X&=&(y_s-x_r \mu_r \mu_s Z^2) X T_{rs},\label{eq:Trscond}\\
      \mu_r S_{rs}(Z^2\ot \id)(x_r-y_s \chi)  &=&\mu_s (Z^2\ot \id) (x_s -y_r \chi) S_{rs}
      \label{eq:Srscond} \eea
      These two conditions are then satisfied if the recursion relations (\ref{eq:WRR}) hold.
      
\end{proof}
It is useful to also consider discrete Fourier transforms of the coefficients $\widehat{W}_{rs}(n)$ and $\overline{W}_{rs}(n)$. Defining \bea W_{rs}(n)=\sli_{m=0}^{N-1}\widehat{W}_{rs}(m) q^{-2 m n}, \quad 
 \widecheck{W}_{rs}(n)=\sli_{m=0}^{N-1}\widebar{W}_{rs}(m) q^{2 m n}\label{eq:WFT},\eea
and using basic results about discrete Fourier transforms, the corresponding recursion relations become
\bea \frac{W_{rs}(n)}{W_{rs}(n-1)}&=& \frac{\mu_r}{\mu_s} \left(\frac{ y_s-x_r q^{2n}}{y_r-x_s q^{2n}}\right),\label{eq:WFTRR1}\\
  \frac{\widecheck{W}_{rs}(n)}{\widecheck{W}_{rs}(n-1)}&=&  \frac{ y_s-x_r q^{2n}\mu_r\mu_s}{y_r-x_s q^{2n}\mu_r\mu_s}.\label{eq:WFTRR2}
\eea

\noindent The coefficients $\overline{W}_{rs}(n)$ and $W_{rs}(n)$ are the Boltzmann weights of the chiral Potts model \cite{BPY88}, and Proposition \ref{Rfact} is the algebraic statement of the factorization property of the R-matrix first observed in \cite{BS90}. \footnote{The additional central elements $z_0,z_1$ were included in the definition of $\uqt$ in \cite{DJMM91b}  in order to achieve this factorization into the standard Chiral Potts weights.}  The discrete Fourier transform $\widecheck{W}_{rs}(n)$ will be used in the proof in Appendix \ref{app:Vfusionproof}. We fix the normalization of our weights by the choice
\bea \widecheck{W}_{rs}(0)=\widehat{W}_{rs}(0)=1.\label{eq:Wnorm}\eea
This normalization will determine the normalization of the coefficients $E_i$ that appear later in Proposition \ref{prop:Vfusion}.

To help picture the various intertwiners, and for later usage it,  we introduce a graphical notation for $S_{rs}$, $T_{rs}$ and $\check{R}_{rr';ss'}$. Representing $\Omega_{rs}$ by 
\vspace{5mm}

\begin{center}
\begin{tikzpicture}[scale=0.4]
\draw[blue,thick] (0,0)--(2,0);
\draw[red,thick] (0,-0.3)--(2,-0.3);
\draw (0,0.15) node[left] {$s\,$};
\draw (0,-0.5) node[left] {$r\,$};
\draw (-2,0) node[left] {$\Omega_{rs}\sim $};
\end{tikzpicture}\hspace{2mm}
\end{center}
we then use the following pictorial representations (with operators acting from West to East and North to South):

\begin{tikzpicture}[scale=0.4]
\Smatrix(0,0);
\draw (-2,0.2) node[left] {$r'$};
\draw (-2,-0.4) node[left] {$r\,$};
\draw (2,0.5) node[right] {$s'$};
\draw (2,-0.1) node[right] {$r'$};
\draw (-0.2,2) node[above] {$s$};
\draw (0.5,2) node[above] {$s'$};
\draw (-0.5,-2) node[below] {$r$};
\draw (0.2,-2) node[below] {$s$};
\draw (-3.5,0) node[left] {$S_{r's}\sim$};
\draw (4,0) node[right] {$:\Omega_{rr'}\ot \Omega_{ss'}\rightarrow \Omega_{rs}\ot  \Omega_{r's'} $};
\end{tikzpicture}
\vspace*{5mm}

\begin{tikzpicture}[scale=0.4]
\Tmatrix(0,0);
\draw (-2,0.2) node[left] {$s$};
\draw (-2,-0.4) node[left] {$r$};
\draw (2,0.2) node[right] {$r$};
\draw (2,-0.4) node[right] {$s$};
\draw (-3.5,0) node[left] {$T_{rs}\sim$};
\draw (4,0) node[right] {$:\Omega_{rs}\rightarrow \Omega_{sr} $};
\end{tikzpicture}

\vspace*{5mm}
\noindent We also define composite operators 
\bea
\check{B}_{r';ss'} &=&(\id \ot T_{r's'}) S_{r's}:\Omega_{rr'} \ot \Omega_{ss'}\rightarrow \Omega_{rs} \ot \Omega_{s'r'}\nonumber\\
\check{A}_{rr';s}&=&S_{rs}( T_{rr'}\ot \id):\Omega_{rr'}\ot \Omega_{ss'}\rightarrow. \Omega_{r's}\ot \Omega_{rs'}
\label{eq:ABdef}\eea
that we shall make use of in Section \ref{sec:TQ}. Their graphical realization follows from that of $S$ and $T$ and is
\vspace*{1mm}

\begin{tikzpicture}[scale=0.4]
\Vmatrix(0,0); 
\draw (-3.5,0) node[left] {$\check{B}_{r';ss'}=(\id\ot T_{r's'}) S_{r's}=$};
\draw (4,0) node[right] {$:\Omega_{rr'}\ot \Omega_{ss'}\rightarrow \Omega_{rs}\ot  \Omega_{s'r'} $};
\draw (-2,-0.4) node[left] {$r\,$};
\draw (-2,0.3) node[left] {$r'$};
\draw (-0.2,2) node[above] {$s$};
\draw (0.5,2) node[above] {$s'$};
\draw (-0.5,-2) node[below] {$r$};
\draw (0.2,-2) node[below] {$s$};
\draw (2,-0.4) node[right] {$s'$};
\draw (2,0.3) node[right] {$r'$};
\end{tikzpicture}
\vspace*{3mm}

\begin{tikzpicture}[scale=0.4]
\Umatrix(0,0); 
\draw (-2,0.6) node[left] {$r'$};
\draw (-2,-0.2) node[left] {$r\,$};
\draw (2,0.6) node[right] {$s'$};
\draw (2,-0.2) node[right] {$r$};
\draw (-0.2,2) node[above] {$s$};
\draw (0.5,2) node[above] {$s'$};
\draw (-0.4,-1.9) node[below] {$r'$};
\draw (0.2,-2.2) node[below] {$s$};
\draw (-3.5,0) node[left] {$\check{A}_{rr';s}=S_{rs}( T_{rr'}\ot \id) =$};
\draw (4,0) node[right] {$:\Omega_{rr'}\ot \Omega_{ss'}\rightarrow  \Omega_{r's}\ot \Omega_{rs'} $};
\end{tikzpicture}

\vspace*{2mm}
\noindent Hence we arrive at the following picture for the operator $\check{R}(rr';ss')$:
\bea
\begin{tikzpicture}[scale=0.4]
 \foreach \r in {0}{
\Vmatrix(3*\r,0);
\Umatrix(3*\r+2.7,-3);
}
\lcorner(0,-2.7);
\rcorner(2.7,-0.3);
\draw (-2,0.2) node[left] {$r'$};
\draw (-2,-0.5) node[left] {$r\,$};

\draw (5,-2.5) node[right] {$r'$};
\draw (5,-3.2) node[right] {$r$};

\draw (-0.2,2) node[above] {$s$};
\draw (0.5,2) node[above] {$s'$};

\draw (2.3,-5.1) node[below] {$s$};
\draw (2.9,-4.8) node[below] {$s'$};

\draw (-3.5,-1.5) node[left] {$\check{R}({rr';ss'})=\check{A}_{rs;s'r'}\circ  \check{B}_{r';ss'}=$};
\draw (6.5,-1.5) node[right] {$:\Omega_{rr'}\ot \Omega_{ss'} \rightarrow \Omega_{ss'}\ot \Omega_{rr'} $};
\end{tikzpicture}
\label{eq:RABfact}\eea

\noindent 
 
Readers familiar with the literature on the chiral Potts model may be wondering why we do not separate the close neighbour pairs of lines, use a picture for $R(rr';ss)$ of the form 
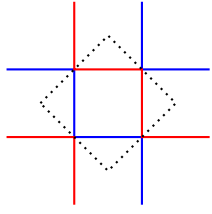
\vspace*{5mm}

\begin{center}
\begin{tikzpicture}[scale=0.3]
\draw[blue,thick] (-3,0)--(0,0) --(0,-3) -- (3,-3) -- (3,-6);
\draw[blue,thick] (3,3) -- (3,0) -- (6,0);
\draw[red,thick] (0,3)--(0,0) --(3,0) -- (3,-3) -- (6,-3);
\draw[red,thick] (-3,-3)--(0,-3)--(0,-6);
\draw[dotted,thick] (-1.5,-1.5)--(1.5,1.5) --(4.5,-1.5)--(1.5,-4.5)--(-1.5,-1.5);
\end{tikzpicture}
\end{center}
\vspace*{5mm}

\noindent and then identify this picture directly with chiral Potts weights along the diagonal lines in the `standard' way (see \cite{BPY88,BS90}). The answer is two fold: (i) separating the red and blue lines might suggest that $\Omega_{rs}$ directly factorizes into the tensor product of two representations, whereas the situation is more subtle as we shall see in the next section; (ii) the standard pictorial identification with diagonal chiral Potts weights $W_{rs}(n)$ and $\wb{W}_{rs}(n)$ corresponds to choosing 
$X$ as a diagonal matrix $X_{ij}= \delta_{i,j}q^{-2i}$ and $Z^2$ as a step operator $(Z^2)_{ij}=\delta_{i,\hbox{mod}(j+1,N)}$.
This is allowed, but it is an unnatural choice from a representation theory point of view in that it would result in $\Omega_{rs}(t_i)$ acting as a step operator. More natural is the choice 
$X_{ij}=\delta_{i,\hbox{mod}(j+1,N)}$, $Z_{ij}= \delta_{i,j}q^i$, for $i,j\in\{0,1,\cdots,N-1\}$. However, for now we make neither choice and leave $X$ and $Z$ as arbitrary $N\times N$ matrices satisfying (\ref{eq:ZYrelns}). 

\subsection{Cyclic Representations of $\uqb$}\label{sec:repnspart2}
Let us define the upper Borel subalgebra $U_q(\mathfrak{b}_+)$ to be the subalgebra of $\uqt$ generated by $e_i,t_i^{\pm 1},z_i$, $i\in\{0,1\}$. Then, following the spirit of the approach of [BLZ97] (which holds for for  the generic-$q$ case with infinite-dimensional representations), we define two new cyclic Borel subalgebra representations $\rho_{r},\bar{\rho}_r:U_q(\mathfrak{b}_+)\rightarrow \End(W)$ that each depend on a single point $r\in\cC_k$:
\ben &&\rho_r (e_0)= -\frac{ \kappa_0 x_r}{q-q^{-1}} X^{-1} ,\quad \rho_r(e_1) = -\frac{\kappa_1 y_r}{q-q^{-1}} X, \quad  \rho_r(t_0)=(q \mu_r)^{-1}Z^{-2}, \quad \rho_r (t_1)=q \mu_r Z^2,\\&& \rho_r(z_0)=\rho_r(z_1)=1,\\[3mm]
&&\bar{\rho}_r(e_0)=  \frac{\kappa_0 y_r}{\mu_r(q-q^{-1})} X^{-1} ,\quad \bar{\rho}_r(e_1) = \frac{\kappa_1 x_r\mu_r}{q-q^{-1}}  X, \quad  \bar{\rho}_r(t_0)=\mu_r^{-1}Z^{-2}, \quad \bar{\rho}_r (t_1)=\mu_r Z^2,\\&& \bar{\rho}_r(z_0)=\bar{\rho}_r(z_1)=1.
\een
\begin{prop}
$\rho_r$ and $\bar{\rho}_r$ are $\uqb$ representations. 
\end{prop}
\begin{proof}
To establish this one needs to check compatibility with the $\uqb$ relations $t_i e_i t_i^{-1} =q^2 e_i$, $t_{1-i} e_i t_{1-i}^{-1} =q^{-2} e_i$ and 
$[e_i,[e_i,[e_i,[e_{1-i}]_{q^{2}}]_1]_{q^{-2}}=0$, for $i\in\{0,1\}$ (where $[a,b]_x=ab-xba$). This compatibilty follows directly from the $ZX=qXZ$ and $[X,X]=0$.
\end{proof}

We need one final,  $U_q(\mathfrak{b}_+)$ representation $\varphi_c:U_q(\mathfrak{b}_+)\rightarrow \End(W)$, $c\in\C^\times$,  defined by
\ben \varphi_c (e_0)=\varphi_c(e_1)=0,\quad \varphi_c(t_0)=\frac{c}{q} Z^{-2},\quad \varphi_c(t_1)=\frac{q}{c} Z^{2},\quad \varphi_c(z_0)=\frac{1}{c},\quad \varphi_c(z_1)=c.
\een
Clearly, $\varphi_c$ is the direct sum of $N$ one-dimensional  $U_q(\mathfrak{b}_+)$ representations.

It is useful to extend our graphics to these new representations in order to have a more intuitive understanding of the results coming in Sections \ref{sec:f&f} and \ref{sec:TQ}.
To this end, let us representing $\rho_r,\bar{\rho}_r,\varphi_c$ by 
\vspace{4mm}

\begin{center}
\begin{tikzpicture}[scale=0.5]
\begin{scope}[shift={(0,0)}]
\draw[red,thick] (0,0)--(2,0);
\draw (0,0) node[left] {$r\,$};
\draw (-1.5,0) node[left] {$\rho_{r}\sim $};
\end{scope}
\begin{scope}[shift={(8,0)}]
\draw[blue,thick] (0,0)--(2,0);
\draw (0,0) node[left] {$r\,$};
\draw (-1.5,0) node[left] {$\bar{\rho}_{r}\sim $};
\end{scope}
\begin{scope}[shift={(16,0)}]
\draw[thick,dashed] (0,0)--(2,0);
\draw (-1.5,0) node[left] {$\varphi_c\sim $};
\end{scope}
\end{tikzpicture}\hspace{2mm}
\end{center}

\begin{prop}
The $U_q(\mathfrak{b}_+)$ representations $\rho_{r},\bar{\rho}_r$ are not restrictions of $\uqt$ representations. 
\end{prop}
\begin{proof} Let us proceed by contradiction. We suppose there exists a $\uqt$ representation $\pi:\uqt\rightarrow \End(W)$ that has the restriction 
\ben \pi(e_0)= \varphi_0 X^{-1},\quad \pi(e_1)= \varphi_1 X,\quad \pi(t_0) = \psi^{-1} Z^{-2},\pi(t_1) = \psi Z^2,
\een
for $\varphi_i$, $\psi\in\C^\times$. Then the relations  $[e_i,f_j]=0$ for $i\neq j$ imply that 
\bea X \pi(f_i)=\pi(f_i) X, \quad i\in\{0,1\}.\label{eq:Xf}
\eea 
It is now useful to consider the two $U_q(\mathfrak{sl}_2)$ Casimir elements $C_i:= q t_i + q^{-1} t_i^{-1} + (q-q^{-1})^2 f_i e_i$.  We have
\begin{eqnarray*}
\pi(C_0) &=& q \psi^{-1} Z^{-2} + q^{-1} \psi Z^2 + (q-q^{-1})^2 \varphi_0 \, \pi(f_0) X^{-1},\\
\pi(C_1) &=& q \psi Z^{2} + q^{-1} \psi^{-1} Z^{-2} + (q-q^{-1})^2 \varphi_1 \, \pi(f_1) X,
\end{eqnarray*}
and hence
\begin{eqnarray}
\pi(f_0) &=& \frac{\left( \pi(C_0) - q \psi^{-1} Z^{-2} - q^{-1} \psi Z^2 \right) X } { (q-q^{-1})^2 \varphi_0 },\label{eq:f0beh}\\
\pi(f_1) &=& \frac{\left( \pi(C_1) - q \psi Z^2- q^{-1} \psi^{-1} Z^{-2} \right) X^{-1} } { (q-q^{-1})^2 \varphi_1 }\label{eq:f1beh}.
\end{eqnarray}
The fact that $[\pi(e_i),\pi(C_i)]=0$ implies that  $[X,\pi(C_i)]=0$. Using the fact that $ZX=qXZ$ , we see that the relations (\ref{eq:f0beh}), (\ref{eq:f1beh}) are incompatible with (\ref{eq:Xf}). 

\end{proof}

\section{Factorization and Fusion of Cyclic $\uqb$ Representation}\label{sec:f&f}
In this section we present a number of intertwiners associated with the representations introduced in Section \ref{sec:repns}. We will make use of these intertwiners in Section \ref{sec:TQ}.
\subsection{Factorization}
We first define an operator $\cO(\chi)\in \hbox{End}(W\ot W)$ as follows:
\begin{defn}\label{def:OP}
    Let $\chi=X^{-1}\ot X$, and let $\cO(\chi)=\sum\limits_{n=0}^{N-1} a_n \chi^n$ 
    be a polynomial with $a_n\in \C$ that satisfies the condition 
    \bea \cO(q \chi)&=&\chi\, \cO(q^{-1} \chi).\label{eq:chi}
    \eea
This condition determines $\cO(\chi)$ up to a multiplicative constant, which we fix by the choice $a_n=q^{-n^2}$.
\end{defn}
\begin{lemma} \label{lem:Oinverse} The operator $\cO(\chi)$ has an inverse of the form $\cO^{-1}(\chi)=\sum\limits_{n=0}^{N-1} b_n \chi^n$.
\end{lemma}

\noindent The proof and form of the $b_n$ is given in Appendix \ref{app:Oinvproof}.
\\

\vspace*{2mm}
The representation $\Omega_{rs}$ defined in Section \ref{sec:Omegadef} is also of course a representation of the Borel subalgebra $\uqb$ and a key result of the current paper is the following Theorem \ref{prop:factor}. We call this result {\it factorization}. 
\begin{thrm} \label{prop:factor} 
We have the following $\uqb$ isomorphism 
\ben \cO(\chi): \Omega_{rs}\ot \varphi_{c_0} \rightarrow \rho_{r}\ot \bar{\rho}_s,\een
where $c_0$ is given in Equation (\ref{eq:c0}).
\end{thrm}
\begin{proof}
We first need to check the statement that 
\ben \cO(\chi) (\Omega_{rs} \ot \varphi_{c_0}) (\Delta(x))= (\rho_{r} \ot \bar{\rho}_{s}) (\Delta(x))\cO(\chi),\quad \hbox{for all  }  x\in \uqb.\een
For $x=z_i$ the equality is clear, and for $x=t_i$ the relation follows from the property $[\chi, Z\ot Z]=0$ and the expression for $c_0$. For $x=e_0$, the relation becomes
 \ben  \frac{\kappa_0 x_r}{q-q^{-1}}  \cO(\chi) \left(X^{-1} \left( \frac{y_s}{x_r \mu_r \mu_s}  Z^{-2} -1\right)\otimes 1 \right) = \frac{ \kappa_0}{q-q^{-1}}( - x_r X^{-1} \ot 1 +\frac{y_s } {q \mu_r \mu_s} Z^{-2} \otimes X^{-1})\cO(\chi)
 \een
 which reduces to 
 \ben q \, \cO(\chi) (X^{-1} Z^{-2} \ot 1) = (Z^{-2}\ot X^{-1}) \cO(\chi).\een
 This in turn becomes
  $\cO(\chi) \, q \chi= \cO(q^2 \chi)$,
 which follows from the defining property (\ref{eq:chi}). Finally, for $x=e_1$ the relations becomes 
 \ben 
 \frac{\kappa_1 y_r}{q-q^{-1}}  \cO(\chi) \left(\left(\frac{x_s\mu_r\mu_s}{y_r} Z^2-1\right)X \otimes 1\right) =\frac{ \kappa_1}{q-q^{-1}}
\left( -y_r X\ot 1 + q  x_s \mu_r \mu_s Z^2 \otimes X\right) \cO(\chi)
 ,\een
 which reduces to \ben \cO(\chi) (Z^2 X\otimes 1)= q (Z^2 \otimes X) \cO(\chi),\een
 which again follows from (\ref{eq:chi}).
 The fact that $\cO(\chi)$ is an isomorphism then follows from Lemma \ref{lem:Oinverse}.
\end{proof}

Again, it is useful to represent the operators $\cO$ and $\cO^{-1}$ and the associated isomorphisms graphically. We do this in the following way:
\begin{center}
\begin{tikzpicture}[scale=0.4]
\Oiso(0,0);
\draw (-2,-0.9) node[left] {$r$};
\draw (-2,-0.3) node[left] {$s$};
\draw (2,-0.9) node[right] {$r$};
\draw (2,1) node[right] {$s$};
\draw (-3,0) node[left] {$\cO\sim$};
\draw (6,0) node[right] {$:\Omega_{rs}\ot \varphi_{c_0} \rightarrow \rho_r\ot \bar{\rho}_s$};
\begin{scope}[shift={(0,-4)}]
\Oisoinv(0,0);
\draw (-2,-0.9) node[left] {$r$};
\draw (-2,1) node[left] {$s$};
\draw (2,-0.9) node[right] {$r$};
\draw (2,-0.3) node[right] {$s$};
\draw (-3,0) node[left] {$\cO^{-1}\sim$};
\draw (6,0) node[right] {$: \rho_r\ot \bar{\rho}_s \rightarrow \Omega_{rs}\ot \varphi_{c_0} $};
\end{scope}
\end{tikzpicture}
\end{center}

\begin{rmk}
    We regard Proposition \ref{prop:factor} as a root-of-unity, cyclic version of the analogous generic-$q$ theorem relating the Verma module to the tensor product of two infinite-dimensional Borel subalgebra representations. This latter result is at the heart of the Q-operator construction and can be found in varying forms throughout the literature including in 
    \cite{BLZ97,BLMS10,BJMST09,BGKNR14}. The statement of this result most resembling the form we have given can be found in \cite{KT14}. Apart from the finite-dimensionality, one other  difference between our root of unity case and the generic-$q$ case is that our $\varphi_c$ representation is a direct sum of 1-dimensional representations (both $e_0$ and $e_1$ act as zero), whereas in the generic-$q$ case its analog is a triangular representation (with one of $e_0$ or $e_1$ acting as zero and the other as a step operator).
\end{rmk}

Before going on to discuss short exact sequences associated with our different cyclic $\uqb$ representations, let us show how $\rho_r$ and $\bar{\rho}_s$ allow us to give a characterization of $T_{rs}$ and $S_{rs}$, and hence of chiral Potts weights, that is an alternative to the conventional picture of Proposition \ref{Rfact}.

\begin{defn}
    Let $\cP(Z)=\sum\limits_{n=0}^{N-1} p_n Z^{2n}$  be a polynomial that satisfies the condition 
    \bea 
    \cP(Z)&=&Z^2 \cP(q^{-1} Z).\label{eq:Deltacond}\eea
\end{defn}
\begin{lemma} The polynomial $\cP(Z)$ is invertible.
\end{lemma}
\begin{proof}
\noindent This follow a very similar argument to the proof of Lemma \ref{lem:Oinverse} given in Appendix \ref{app:Oinvproof}.
\end{proof}

\begin{prop}\label{prop:STalt}
  Defining $\mathfrak{T}_{rs}= \cO(\chi) (T_{rs}\ot \id) \cO^{-1}(\chi)$, and $\cS_{rs}=(\cP(Z)^{-1}\ot \id) S_{rs} (\cP(Z)\ot \id)$, we have the following $\uqb$ isomorphisms
  
  \ben 
  (i) \quad\mathfrak{T}_{rs}&:& \rho_r\ot \bar{\rho}_s \rightarrow \rho_s \ot \bar{\rho}_r,\\
  (ii)\quad \cS_{rs}&:& \bar{\rho}_r\ot \rho_s \rightarrow \bar{\rho}_s\ot \rho_r.
  \een
  \end{prop}
  \begin{proof}
      Statement $(i)$ follows as a simple corollary of Propositions \ref{Rfact} and \ref{prop:factor}. To prove (ii), we first note that we indeed have
      \ben \cS_{rs} \, (\bar{\rho}_r\ot \rho_s )\Delta(t_i) = (\bar{\rho}_s\ot \rho_r )\Delta(t_i) \, \cS_{rs}.\een
      For commutation with $\Delta(e_0)$, we need 
      \ben \frac{1}{\mu_r}\cS_{rs}\left(-y_r X^{-1}\ot \id +
      x_s Z^{-2}\ot X^{-1}\right)=\frac{1}{\mu_s}\left(-y_s X^{-1}\ot \id +
      x_r Z^{-2}\ot X^{-1}\right)\cS_{rs},\een
      which can be re-expressed using (\ref{eq:Deltacond}) as
      \ben \frac{1}{\mu_r} S_{rs}\left(-y_r X^{-1}Z^{-2}\ot \id +
       x_s Z^{-2}\ot X^{-1}\right)=\frac{1}{\mu_s}\left(- y_s X^{-1}Z^{-2}\ot \id +
       x_r Z^{-2}\ot X^{-1}\right) S_{rs},\een
      which is equivalent to the earlier condition
      (\ref{eq:Srscond}).
      For $\Delta(e_1)$ the requirement for commutation is \ben 
      \mu_r \cS_{rs} \left(-x_r X\ot \id +
      y_s Z^2\ot X\right)=\mu_s \left(-x_s X\ot \id +
      y_r Z^2\ot X\right)S_{rs},
      \een which becomes 
      \ben 
      \mu_r S_{rs} \left(-x_r Z^2 X\ot \id +
      y_s Z^2\ot X\right)=\mu_s \left(-x_s Z^2 X\ot \id +
      y_r Z^2\ot X\right)S_{rs}
      \een
      which can again be re-expressed as the condition (\ref{eq:Srscond}). 
  \end{proof}

\subsection{Fusion}
We now present three short exact sequences (SESs) for different $\uqb$ representations. In analogy with the classic results of \cite{KR81}, we refer to these properties as {\it fusion}. The SESs also involve the standard two-dimensional evaluation representation $\pi_z:\uqt\rightarrow \End{V}$ ($z\in \C^\times$, $V=\C^2$) which we define by

\begin{equation} \label{eq:2Drep}
  \begin{aligned}
    & \pi_{z} (e_1) = \begin{pmatrix}
        0&z\\0&0
    \end{pmatrix}, \quad 
    & \pi_{z} (f_1) = \begin{pmatrix}
        0&0\\z^{-1}&0
        \end{pmatrix}, \quad
 & \pi_{z} (t_1) = \begin{pmatrix}
        q&0\\0&q^{-1}
        \end{pmatrix},   \quad \pi_{z} (z_1)=1,   
     \\
    & \pi_{z} (e_0) = \begin{pmatrix}
        0&0\\z&0
    \end{pmatrix}, \quad 
    & \pi_{z} (f_0) = \begin{pmatrix}
        0&z^{-1}\\ 0&0
        \end{pmatrix}, \quad
 & \pi_{z} (t_0) = \begin{pmatrix}
        q^{-1}&0\\0&q
        \end{pmatrix},   \quad \pi_{z} (z_0)=1.   
    \end{aligned} 
\end{equation}
Then we have the following:
\begin{thrm}\label{prop:SESs} Let $r,s\in\cC_k$, with $s=(x_s,y_s,\mu_s)$. Defining  $sq^{\pm 1}=(x_s q^{\pm 1},y_s q^{\pm 1},\mu_s q^{\pm 1})$, we see that we also have $s q^{\pm 1}\in \cC_k$. Choose $z_s$ to satisfy
     $z_s^2=\kappa_0 \kappa_1 x_s y_s $ . Let us define the following complex constants
     \ben c_s=-\frac{\kappa_0 x_s \mu_s q^2 }{z_s},\quad \bar{c}_s=\frac{\kappa_0 y_s q}{z_s},\quad d_s=\frac{\kappa_0 y_s}{z_s q}.\een   
     Then we have the following $\uqb$ short exact sequences:
\begin{equation*}
\begin{aligned}
&(i) &
\begin{tikzcd}[column sep=scriptsize]
0 \arrow[r]  & \rho_{sq} \arrow[r,"\iota_s"] & \rho_{s} \ot \pi_{z_s} \arrow[r,"\tau_s"]  & \rho_{sq^{-1}} \arrow[r] & 0 
\end{tikzcd}\;
&\hbox{where}\quad\iota_s=\begin{pmatrix} c_s Z\\X Z^{-1} \end{pmatrix},\; \tau_s=(-\frac{q }{c_s} XZ^{-1}, Z),\\
&(ii) &
\begin{tikzcd}[column sep=scriptsize]
0 \arrow[r] & \bar{\rho}_{sq} \arrow[r,"\bar{\iota}_s"] & \bar{\rho}_{s} \ot \pi_{z_s} \arrow[r,"\bar{\tau}_s"]  & \bar{\rho}_{sq^{-1}} \arrow[r] & 0 
\end{tikzcd}\;
&\hbox{where} \quad\bar{\iota}_s=\begin{pmatrix} \bar{c}_s \\X Z^{-2} \end{pmatrix},\; \bar{\tau_s}=(-\frac{1 }{\bar{c}_s} XZ^{-2}, 1),\\
&(iii) &\hspace*{-2mm}
\begin{tikzcd}[column sep=scriptsize]
0 \arrow[r] & \Omega_{r,sq} \arrow[r,"\bar{I}_s"] & \Omega_{r,s} \ot \pi_{z_s} \arrow[r,"\bar{T}_s"]  & \Omega_{r,sq^{-1}} \arrow[r] & 0 
\end{tikzcd}\;
&\hbox{where} \quad \bar{I}_s=\begin{pmatrix} d_s\\X\end{pmatrix},\; \bar{T}_s=(-X,d_s).
\end{aligned}
\end{equation*}
\end{thrm}
\begin{proof}
Let us outline the proof of (i). The proofs of (ii) and (iii) are then very similar. First of all, the required $\uqb$ intertwining properties are equivalent to \ben 
\quad (\rho_s\ot \pi_{z_s}) (\Delta(x)) \circ \iota_s = \iota_s \circ \rho_{sq} (x) , \quad
\tau_s \circ (\rho_s\ot \pi_{z_s}) (\Delta(x))=  \rho_{sq^{-1}} \circ \tau_s,
\quad \forall x\in \uqb.\een
These relations are just checked by explicit computation. The injectivity and surjectivity are of course equivalent to Ker$(\iota_s)$=0, Im($\tau_s$)=$W$. These relations are also simple to check. Finally the SES property then follows from the fact that $\tau_s\circ \iota_s=-q  X  + ZXZ^{-1}=0$.

\end{proof}
\begin{rmk} We note that while $\Omega_{rs}$ and $\pi_z$ appearing in $(iii)$ are also full $\uqt$ representations (unlike the $\rho_{s}$ and $\bar{\rho}_s$ representations of $(i)$ and $(ii)$), $\bar{I}_s$ and $\bar{T}_s$ are only $\uqb$ intertwiners and are {\it not} $\uqt$ intertwiners. A similar short exact sequence for cyclic representations of the full $\uq$ algebra was introduced and studied in \cite{Korff_2003a,Korff_2003b}. 
\end{rmk}
Returning to the development of our graphical dictionary, we introduce the new pictorial representation \vspace*{3mm}

\begin{tikzpicture}[scale=0.5]
\draw[green,thick] (0,0)--(2,0);
\draw (0,0) node[left] {$z\,$};
\draw (-1.5,0) node[left] {$\pi_z\sim $};
\end{tikzpicture}

\noindent and then represent the injections and surjections involved in Proposition \ref{prop:SESs} as \vspace*{5mm}

\begin{tikzpicture}[scale=0.5]
\begin{scope}[shift={(0,0)}]
\draw[red,thick] (0,0)--(2,0);
\draw[thick,green,rounded corners=12] (1,0) -- (1,1) -- (2,1);
\draw (0,0) node[left] {$sq\,$};
\draw (2,0) node[right] {$s\,$};
\draw (2,1) node[right] {$z_s\,$};
\draw (-1.5,0) node[left] {${\iota}_s\sim $};
\draw (3,0) node[right] {$:{\rho}_{sq}\rightarrow {\rho}_s\ot \pi_{z_s}$};
\end{scope}
\begin{scope}[shift={(16,0)}]
\draw[red,thick] (0,0)--(2,0);
\draw[thick,green,rounded corners=12] (0,1) -- (1,1) -- (1,0);
\draw (0,0) node[left] {$s\,$};
\draw (2,0) node[right] {$sq^{-1}\,$};
\draw (0,1) node[left] {$z_s\,$};
\draw (-1.5,0) node[left] {${\tau}_s\sim $};
\draw (4,0) node[right] {$:{\rho}_{s}\ot \pi_{z_s}\rightarrow {\rho}_{sq^{-1}}$};
\end{scope}
\begin{scope}[shift={(0,-3.5)}]
\draw[blue,thick] (0,0)--(2,0);
\draw[thick,green,rounded corners=12] (1,0) -- (1,1) -- (2,1);
\draw (0,0) node[left] {$sq\,$};
\draw (2,0) node[right] {$s\,$};
\draw (2,1) node[right] {$z_s\,$};
\draw (-1.5,0) node[left] {$\bar{\iota}_s\sim $};
\draw (3,0) node[right] {$:\bar{\rho}_{sq}\rightarrow \bar{\rho}_s\ot \pi_{z_s}$};
\end{scope}
\begin{scope}[shift={(16,-3.5)}]
\draw[blue,thick] (0,0)--(2,0);
\draw[thick,green,rounded corners=12] (0,1) -- (1,1) -- (1,0);
\draw (0,0) node[left] {$s\,$};
\draw (2,0) node[right] {$sq^{-1}\,$};
\draw (0,1) node[left] {$z_s\,$};
\draw (-1.5,0) node[left] {$\bar{\tau}_s\sim $};
\draw (4,0) node[right] {$:\bar{\rho}_{s}\ot \pi_{z_s}\rightarrow \bar{\rho}_{sq^{-1}}$};
\end{scope}
\begin{scope}[shift={(0,-7)}]
\draw[blue,thick] (0,0)--(2,0);
\draw[red,thick] (0,-0.3)--(2,-0.3);
\draw[thick,green,rounded corners=12] (1,0) -- (1,1) -- (2,1);
\draw (0,0.2) node[left] {$sq\,$};
\draw (2,0.2) node[right] {$s\,$};
\draw (0,-0.5) node[left] {$r\,$};
\draw (2,-0.5) node[right] {$r\,$};
\draw (2,1) node[right] {$z_s\,$};
\draw (-1.5,0) node[left] {$\bar{I}_s\sim $};
\draw (3,0) node[right] {$:\Omega_{r,sq}\rightarrow \Omega_{r,s}\ot \pi_{z_s}$};
\end{scope}
\begin{scope}[shift={(16,-7)}]
\draw[blue,thick] (0,0)--(2,0);
\draw[red,thick] (0,-0.3)--(2,-0.3);
\draw[thick,green,rounded corners=12] (0,1) -- (1,1) -- (1,0);
\draw (0,0.2) node[left] {$s\,$};
\draw (2,0.2) node[right] {$sq^{-1}\,$};
\draw (0,-0.5) node[left] {$r\,$};
\draw (2,-0.5) node[right] {$r\,$};
\draw (0,1) node[left] {$z_s\,$};
\draw (-1.5,0) node[left] {$\bar{T}_s\sim $};
\draw (4,0) node[right] {$:\Omega_{r,s}\ot \pi_{z_s}\rightarrow \Omega_{r,sq^{-1}}$};
\end{scope}
\end{tikzpicture}

\subsection{Factorization and Fusion of L-operators}\label{sec:Lops}
The factorization and fusion relations presented above also manifest themselves in terms of L-operators, and historically this is the form in which such relations have usually been discovered (see for example \cite{BS90,BLMS10}). The L-operators are also essential to consider as they are the building blocks of the T and Q operators defined in the Section \ref{sec:TQ}.

If $\Pi:\uqb\rightarrow \End(W)$ denotes one of the four cyclic $\uqb$ representation defined in Section 2 (i.e., $\Pi$ is one of $\Omega_{rs},\varphi_c,\rho_r,\bar{\rho}_r$), then the corresponding L-operator $\check{L}_{\Pi}(z):W\ot V\rightarrow V\ot W$ is defined up to a normalization factor by the requirement \bea  \check{L}_\Pi(z) (\Pi\ot \pi_z) \Delta(x) = (\pi_z\ot \Pi) \Delta(x) \check{L}_\Pi(z)\quad \hbox {for all   }x\in \uqb. \label{eq:Lhom}\eea
In the case of $\check{L}_{\Omega_{rs}}(z)$, the relation (\ref{eq:Lhom}) also extends to all $x\in\uqt$.
As $\Pi\ot \pi_z$ is generically irreducible as a $\uqb$ representation, this requirement fixes $\check{L}_{\Pi}(z)$ uniquely up to an arbitrary multiplicative constant. As is usual when discussing R-matrices and L-operators, it is also useful for us to consider the associated operator
\bea L_{\Pi}(z)= P \check{L}_\Pi(z), \quad \hbox{where}\quad  P(a\ot b) = b\ot a.\eea
Clearly we then have that $L_{\Pi}(z)\in\End(W\ot V)$ with
\bea  {L}_\Pi(z) (\Pi\ot \pi_z) \Delta(x) = (\Pi \ot \pi_z) \Delta^{op}(x) {L}_\Pi(z),\quad \hbox {for all   }x\in \uqb. \label{eq:Lhom2}\eea
The `rule' that we shall following for deciding which of $\check{L}_{\Pi}(z)$ or $L_{\Pi}(z)$ to use is that $\check{L}_{\Pi}$ is more natural when stating algebraic results and identifying with pictures, and $L_{\Pi}(z)$ easier for defining transfer matrices or Q-operators.

One other advantage of the ${L}_{\Pi}(z)$ version is that, since it is an element of 
$\End(W\ot V)$, we can represent it as a $2\times 2$ matrix with entries in $\End(W)$.
In fact, all four ${L}_\Pi(z)$ have a common form. 
\noindent In order to express this form succinctly it is useful to introduce the following notation. Suppose $A$ and $B$ are $2\times 2$ matrices with entries in $\mathbb{C}$. Then define $\{A,B\}\in \End(W\ot V)$ as follows:
\bea \{A,B\}=\begin{pmatrix}
X^{-1} & 0\\0&1
\end{pmatrix}
A
\begin{pmatrix}
Z^{-1} & 0\\0&Z
\end{pmatrix}
B
\begin{pmatrix}
X & 0\\0&1
\end{pmatrix}.
\label{eq:ABnotn}\eea
Now let us define two matrices that depend on a point $r=(x,y,\mu)\in \cC_k$ and $z\in \mathbb{C}^\times$ (with $\kappa_i$ the same constants as in Section 2):
\bea
U_r(z)= 
\begin{pmatrix}
    z&  \kappa_0 x \mu\\ \kappa_1 y&  z \mu
\end{pmatrix},\quad 
V_r(z)= 
\begin{pmatrix}
    -q z& \kappa_0 y \\ 
    q\kappa_1 x \mu & -z\mu
\end{pmatrix}.
\label{eq:UVdef}\eea
Using the notation (\ref{eq:ABnotn}), and after choosing a normalization, the four L-operators are given by
\bea
{L}_{\Omega_{rs}}(z)&=&\{U_r(z),V_s(z)\},\quad {L}_{\varphi_c}(z)=\{\id,\id\},\nonumber\\
{L}_{\rho_{r}}(z)&=&\{U_r(z),\id\},\quad {L}_{\bar{\rho}_r}(z)=\{V_r(z),\id\},
\label{eq:Ldefs}\eea
where $\id$ is the 2$\times$2 unit matrix. As ${L}_{\varphi_c}(z)$ is idependent of both $z$ and the constant $c$, we shall henceforth denote it as simply $L_\varphi$.
\rmk 
Explicit expressions for ${L}_{\Omega_{rs}}$
are not new - they appeared in \cite{BS90} and have been used by various authors, for example in
\cite{Korff_2003a,Niccoli2010,Maillet2018}.
The standard form (\ref{eq:Ldefs}) of L-operators that we use was inspired by the paper \cite{MLP21} which introduces a similar form for an L-operator analogous to ${L}_{\Omega_{rs}}$ but for the case of {\it semi-cyclic} representations. The latter are discussed briefly in Section \ref{sec:discussion}.

\rmkend \vspace*{2mm}

In terms of the graphical notation we have already introduced, the four L-operators are given by the obvious pictures:\vspace*{5mm}

\begin{tikzpicture}[scale=0.4]
\begin{scope}[shift={(0,0)}]
\LOmega(0,0,r,s,z);
\draw(-3,0) node[left] {$\check{L}_{\Omega_{rs}}(z)\sim$};
\draw(2.5,0) node[right] {$:\Omega_{rs}\ot \pi_z\rightarrow \pi_z\ot \Omega_{rs}$};
\end{scope}
\begin{scope}[shift={(22,0)}]
\Ltau(0,0,z);
\draw(-3,0) node[left] {$\check{L}_{\varphi}\sim$};
\draw(2.5,0) node[right] {$:\varphi_c\ot \pi_z\rightarrow \pi_z\ot \varphi_c$};
\end{scope}
\begin{scope}[shift={(0,-6)}]
\Lrho(0,0,r,z);
\draw(-3,0) node[left] {$\check{L}_{\rho_{r}}(z)\sim$};
\draw(2.5,0) node[right] {$:\rho_{r}\ot \pi_z\rightarrow \pi_z\ot \rho_{r}$};
\end{scope}
\begin{scope}[shift={(22,-6)}]
\Lrhobar(0,0,r,z);
\draw(-3,0) node[left] {$\check{L}_{\bar{\rho}_{r}}(z)\sim$};
\draw(2.5,0) node[right] {$:\bar{\rho}_{r}\ot \pi_z\rightarrow \pi_z\ot \bar{\rho}_{r}$};
\end{scope}
\end{tikzpicture}

We will also need one other $\uqt$ isomorphism ${\bf \check{L}}_{\Omega_{rs}}(z):V \ot W \rightarrow W \ot V$ that satisfies 
\bea  \check{\bf L}_{\Omega_{rs}}(z)
(\pi_z\ot \Pi) \Delta(x) = (\Pi\ot \pi_z) \Delta(x)  \check{\bf L}_{\Omega_{rs}}(z)\quad\hbox{for all   }x\in \uqt. \label{eq:Lbfdef}
\eea
We can clearly obtain this by inverting $\check{L}_{\Omega_{r,s}}(z)$, but we fix the normalization differently such that 
\ben \check{\bf L}_{\Omega_{rs}}(z) \check{L}_{\Omega_{r,s}}(z) = \check{L}_{\Omega_{r,s}}(z)\check{\bf L}_{\Omega_{rs}}(z) =q^2 (z^2-z_r^2)(z^2-z_s^2) \mu_r \mu_s  \, \id,\een
where $z_r^2=\kappa_0 \kappa_1 x_r y_r$ as above. The explicit expression for this $\check{\bf L}_{\Omega_{rs}}(z)$ is given in Appendix \ref{app:Vfusionproof}. We represent this operator graphically by
\vspace*{2mm}
\begin{center}
\begin{tikzpicture}[scale=0.4]
\LOmegainv(0,0,z,r,s);
\draw(-4,0) node[left] {$\check{\bf L}_{\Omega_{rs}}(z)\sim$};
\draw(4,0) node[right] {$:\pi_z\ot \Omega_{rs} \rightarrow \Omega_{rs}\ot \pi_z$};
\end{tikzpicture}
\end{center}

Our first result concerns the basic properties of $\check{L}_{\Omega_{r,s}}(z)$ which are the analogues of Proposition \ref{Rfact}.
\begin{prop}\label{prop:LLproperties}
    The L-operators defined by (\ref{eq:Ldefs}) satisfy the following relations:
  \ben
&(i)&\quad(\id \ot  T_{rs}) \, \check{L}_{\Omega_{rs}}(z) =   \check{L}_{\Omega_{sr}}(z)\,  (T_{rs}\ot \id),\\
&(ii)&\quad (\id \ot S_{r's})\, \check{L}_{\Omega_{rr'}}(z) \ot \check{L}_{\Omega_{ss'}}(z) =  \check{L}_{\Omega_{rs}}(z) \ot \check{L}_{\Omega_{r's'}}(z)\, (S_{r's}\ot \id) ,\\
&(iii)&\quad (\id \ot \check{R}_{rr';ss'} )\, \check{L}_{\Omega_{rr'}}(z) \ot \check{L}_{\Omega_{ss'}}(z) = \check{L}_{\Omega_{ss'}}(z) \ot \check{L}_{\Omega_{rr'}}(z) (\check{R}_{rr';ss'}\ot \id).
\een
\end{prop}
\begin{proof}
We could prove each of these relations by using uniqueness of the associated $\uqt$ intertwiner, but instead we take a more direct linear-algebra approach that exploits the simplicity of the form of the L-operators given by \ref{eq:Ldefs} (again, this approach is inspired by \cite{MLP21}).

\vspace*{3mm}

\noindent   (i) Using the notation of (\ref{eq:ABnotn}) and (\ref{eq:UVdef}), the statement (i) is equivalent to 
\ben T_{rs}(Z)\, \{U_r(z),V_s(z)\} = \{U_s(z),V_r(z)\} \,T_{rs}(Z),\een
which reduces to  
\ben \begin{pmatrix}
T_{rs}(Z q^{-1}) & 0\\0&T_{rs}(Z)
\end{pmatrix}
U_r(z)
\begin{pmatrix}
Z^{-1} & 0\\0&Z
\end{pmatrix}
V_s(z)
=
U_s(z)
\begin{pmatrix}
Z^{-1} & 0\\0&Z
\end{pmatrix}
V_r(z)
\begin{pmatrix}
T_{rs}(Z q^{-1}) & 0\\0&T_{rs}(Z)
\end{pmatrix}.
\een
This is equivalent to 
\ben T_{rs}(q^{-1} Z) (y_s-x_r \mu_r \mu_s Z^2) = T_{rs}(Z) (y_r-x_s \mu_r \mu_s Z^2),\een
which is equivalent to the $T_{rs}$ defining relation (\ref{eq:Trscond}).

\vspace*{3mm}
\noindent (ii) The statement is equivalent to 
\ben S_{r's}(\chi) \{U_r(z),V_{r'}(z)\} \ot \{U_s(z),V_{s'}(z)\}  = 
\{U_r(z),V_{s}(z)\} \ot \{U_{r'}(z),V_{s'}(z) \}S_{r's}(\chi),\een
 where $\chi:=X^{-1}\ot X$
 and hence
 \ben &&S_{r's} (\chi)
 \begin{pmatrix}
Z_1^{-1} & 0\\0&Z_1
\end{pmatrix}
V_{r'}(z)
\begin{pmatrix}
\chi^{-1} & 0\\0&1
\end{pmatrix}
U_s(z)
\begin{pmatrix}
Z_2^{-1} & 0\\0&Z_2
\end{pmatrix}
\\
&& =
 \begin{pmatrix}
Z_1^{-1} & 0\\0&Z_1
\end{pmatrix}
V_s(z)
\begin{pmatrix}
\chi^{-1} & 0\\0&1
\end{pmatrix}
U_{r'}(z)
\begin{pmatrix}
Z_2^{-1} & 0\\0&Z_2
\end{pmatrix}
S_{r's},
\een
which simplifies to 
\ben &&
 \begin{pmatrix}
S_{r's}(q^{-1}\chi) & 0\\0& S_{rs'} (q\chi)
\end{pmatrix}
V_{r'}(z)
\begin{pmatrix}
\chi^{-1} & 0\\0&1
\end{pmatrix}
U_s(z)
\\
&& =
V_s(z)
\begin{pmatrix}
\chi^{-1} & 0\\0&1
\end{pmatrix}
U_{r'}(z)
\begin{pmatrix}
 S_{r's}(q^{-1} \chi)& 0\\0& S_{rs'}(q \chi)
\end{pmatrix}.
\een
This is equivalent to the relation
\ben S_{r's}(q^{-1} \chi) \mu_s (q x_s-y_{r'} \chi)= S_{r's}(q\chi) \mu_{r'}(q x_{r'}-y_s \chi),\een
which in turn is equivalent to the defining condition (\ref{eq:Srscond}).

\vspace*{3mm}
\noindent (iii) This statement follows directly from (i) and (ii).
\end{proof}

Let us now introduce the standard 6-vertex model R-matrix defined algebraically as the $\uqt$ isomorphism
\ben \check{R}(z/w): \pi_z\ot \pi_w \rightarrow \pi_w\ot \pi_z.\een
Fixing the arbitrary multiplicative normalization we have\footnote{This may look like we have swapped  $b(z)$ and $c(z)$ compared to usual conventions, but this is simply because we are dealing with $\check{R}=P R$.} 

\bea &\check{R}(z)=\begin{pmatrix}
a(z)&0&0&0\\
0&c(z)&b(z)&0\\
0&b(z)&c(z)&0\\
0&0&0&a(z)
\end{pmatrix},\;\;\;\hbox{with} \;(a(z),b(z),c(z))=\left(1-q^2 z^2, q (1-z^2), z (1-q^2)\right).\nonumber\\ &\label{eq:Rmatrix}\eea

\noindent We use the obvious graphical representation
\begin{center}
\begin{tikzpicture}[scale=0.3]
\draw[green,thick] (-2,0)--(2,0);
\draw[green,thick] (0,2)--(0,-2);
\draw (-2,0) node[left] {$z$};
\draw (0,2) node[above] {$w$};
\draw (-4,0) node[left] {$\check{R}(z/w)\,\sim$};
\end{tikzpicture}
\end{center}

We then have the following result concerning the commutation of L-operators.
\begin{prop}\label{prop:RLL}
As operators on $W\ot V\ot V\rightarrow V\ot V\ot W$ we have the identities:
\ben &&(i)\quad(\check{R}(z/w)\ot \id)(\id\ot \check{L}_{\rho_r}(w)) (\check{L}_{\rho_r}(z)\ot \id)=
(\id\ot \check{L}_{\rho_r}(z)) (\check{L}_{\rho_r}(w)\ot \id)(\check{R}(z/w)\ot \id),\\
&&(ii)\quad(\check{R}(z/w)\ot \id)(\id\ot \check{L}_{\bar{\rho}_r}(w)) (\check{L}_{\bar{\rho}_r}(z)\ot \id)=
(\id\ot \check{L}_{\bar{\rho_r}}(z)) (\check{L}_{\bar{\rho_r}}(w)\ot \id)(\check{R}(z/w)\ot \id).
\een
\end{prop}
\begin{proof}
The proof again is either by uniqueness of the associated intertwiners or by explicit checking of the relations.
\end{proof}
\noindent The pictures corresponding to Proposition \ref{prop:RLL} are

\vspace*{3mm}
 \begin{tikzpicture}[scale=0.3]
 \begin{scope}[shift={(0,0)}]
 \draw[thick,red,rounded corners=5] (-2,-1)--(-0.2,1)--(2.5,1);
 \draw[thick,green,rounded corners=5] (-2,1)--(-0.2,-1)--(2.5,-1);
\draw[thick,green] (1,2.5)--(1,-2.5); 
\draw (-2,1) node[left] {$z$};\draw (-2,-1) node[left] {$r$}; \draw (1,2.5) node[above] {$w$}; \draw(3.5,0) node[] {$=$}; \draw(-3.5,0) node[left] {$(i)$};
\end{scope}
 \begin{scope}[shift={(8,0)}]
 \draw[thick,red,rounded corners=5] (2,1)--(0.2,-1)--(-2.5,-1);
 \draw[thick,green,rounded corners=5] (2,-1)--(0.2,1)--(-2.5,1);
\draw[thick,green] (-1,-2.5)--(-1,2.5); 
\draw (-2.5,-1) node[left] {$r$};\draw (-2.5,1) node[left] {$z$}; \draw (-1,2.5) node[above] {$w$}; 
\end{scope}
 \begin{scope}[shift={(20,0)}]
 \draw[thick,blue,rounded corners=5] (-2,-1)--(-0.2,1)--(2.5,1);
 \draw[thick,green,rounded corners=5] (-2,1)--(-0.2,-1)--(2.5,-1);
\draw[thick,green] (1,2.5)--(1,-2.5); 
\draw (-2,1) node[left] {$z$};\draw (-2,-1) node[left] {$r$}; \draw (1,2.5) node[above] {$w$}; \draw(3.5,0) node[] {$=$}; \draw(-3.5,0) node[left] {$(ii)$};
\end{scope}
 \begin{scope}[shift={(28,0)}]
 \draw[thick,blue,rounded corners=5] (2,1)--(0.2,-1)--(-2.5,-1);
 \draw[thick,green,rounded corners=5] (2,-1)--(0.2,1)--(-2.5,1);
\draw[thick,green] (-1,-2.5)--(-1,2.5); 
\draw (-2.5,-1) node[left] {$r$};\draw (-2.5,1) node[left] {$z$}; \draw (-1,2.5) node[above] {$w$}; 
\end{scope}
\end{tikzpicture}
\vspace*{2mm}

\noindent where graphical composition runs from West to East and North to South as before.

\begin{rmk} 
The approach of \cite{BS90} used L-operators alone: their starting point was to first construct L-operators satisfying $$(\check{R}(z/w)\ot \id)(\id\ot \check{L}_{\Omega_{rs}}(w)) (\check{L}_{\Omega_{rs}}(z)\ot \id)=
(\id\ot \check{L}_{\Omega_{rs}}(z)) (\check{L}_{\Omega_{rs}}(w)\ot \id)(\check{R}(z/w)\ot \id)$$ and then find an R-matrix satisfying property $(iii)$ of Proposition \ref{prop:LLproperties}. This yielded an R-matrix of the factorized form given by our Equation (\ref{eq:Rfact}).
\end{rmk}

\subsubsection{Factorization of L-operators}
In the language of L-operators, we have the following analogue of Proposition \ref{prop:factor}.

\begin{prop}\label{prop:Lproperties}
    The L-operators of Equation (\ref{eq:Ldefs}) satisfy the following factorization relation:
  \ben
&(\id \ot \cO(\chi))\, \check{L}_{\Omega_{rs}}(z)\ot \check{L}_\varphi =   \check{L}_{\rho_{r}}(z)\ot \check{L}_{\bar{\rho}_s}(z)\, (\cO(\chi)\ot \id).
\een
\end{prop}

\noindent Note that the statement in this proposition is represented graphically by

\vspace*{2mm}
\begin{center}
\begin{tikzpicture}[scale=0.4]
\Oiso(0,0);
\draw (-2,-1) node[left] {$r$};
\draw (-2,-0.4) node[left] {$s$};
\draw (2,-1) node[right] {$r$};
\draw (2,1) node[right] {$s$};
\draw[thick,green] (-1,2) -- (-1,-2);
\draw (-1,2) node[above] {$z$};
\Oiso(8,0);
\draw (6,-1) node[left] {$r$};
\draw (6,-0.4) node[left] {$s$};
\draw (10,-1) node[right] {$r$};
\draw (10,1) node[right] {$s$};
\draw[thick,green] (9,2) -- (9,-2);
\draw (9,2) node[above] {$z$};
\draw (4,0) node[] {$=$};
\end{tikzpicture}
\end{center}
\vspace*{1mm}

\begin{proof} This statement can be rewritten as 
 \ben \cO(\chi)\, \{U_r(z),V_s(z)\}\ot \{\id,\id \} &=& \{U_r(z),\id \}\ot \{V_s(z),\id\}\, \cO(\chi)\een
 which follows from the more general property 
 \bea \cO(\chi)\, \{A,B\}\ot \{\id,C\} &=& \{A,\id\}\ot \{B,C\}\,\cO(\chi)\label{eq:ABC} \eea
 where $A,B,C$ are any $N\times N$ complex valued matrices.
 Property (\ref{eq:ABC}) is equivalent to \ben &&
 \begin{pmatrix}
\cO (q^{-1}\chi) & 0\\0&  \cO (q\chi)
\end{pmatrix}
B
\begin{pmatrix}
\chi^{-1} & 0\\0&1
\end{pmatrix}
\\
&& =
\begin{pmatrix}
\chi^{-1} & 0\\0&1
\end{pmatrix}
B
\begin{pmatrix}
\cO(q^{-1} \chi)& 0\\0& \cO(q \chi)
\end{pmatrix}.
\een
 This requirement in turn reduces to
 $\cO(q \chi)=\chi \, \cO(q^{-1} \chi)$, which is the defining property (\ref{eq:chi}) 
 of $\cO(\chi)$.
\end{proof}

\subsubsection{Fusion of L-operators}
Now we turn to the SES relations of Theorem \ref{prop:SESs}, and their manifestation in terms of L-operators. 
\begin{prop}\label{prop:Lfusion}
We have the following equalities
\ben
&&(i)\quad \, (\check{L}_{{\rho}_s}(w)\ot \id)(\id\ot \check{R}(z_s/w))({\iota}_s\ot\id )= C_1(s,w) (\id \ot {\iota}_s) \check{L}_{\rho_{sq}}(w) : W\ot V \rightarrow V\ot W\ot V\\
&&(ii)\quad (\id \ot{\tau}_s) (\check{L}_{{\rho}_{s}}(w)\ot \id)(\id \ot \check{R}(z_s/w)) = C_2(s,w) \, \check{L}_{{\rho}_{sq^{-1}}}(w) ({\tau}_s\ot \id): W\ot V \ot V \rightarrow V\ot W ,\\[3mm]
&&(iii)\quad \, (\check{L}_{\bar{\rho}_s}(w)\ot \id)(\id\ot \check{R}(z_s/w))(\bar{\iota}_s\ot\id )= C_1(s,w) (\id \ot \bar{\iota}_s) \check{L}_{\bar{\rho}_{sq}}(w) : W\ot V \rightarrow V\ot W\ot V\\
&&(iv)\quad (\id \ot\bar{\tau}_s) (\check{L}_{\bar{\rho}_{s}}(w)\ot \id)(\id \ot \check{R}(z_s/w)) = C_2(s,w) \, \check{L}_{\bar{\rho}_{sq^{-1}}}(w) (\bar{\tau}_s\ot \id): W\ot V \ot V \rightarrow V\ot W ,
\een
where the coefficient are given by
\ben C_1(s,w)=q^{-1} b(z_s/w),\quad C_2(s,w)=q a(z_s/w),\een
and $z_s^2=\kappa_0\kappa_1 x_sy_s$ as in Theorem \ref{prop:SESs}.
\end{prop}

\noindent The proof of Proposition \ref{prop:Lfusion} is given in Appendix \ref{app:Lfusionproof}. The meaning of these four relations, which are sometimes referred to as `bootstrap' relations, becomes more transparent when viewing their corresponding graphical realizations:
\vspace*{5mm}

\begin{tikzpicture}[scale=0.4]
\begin{scope}[shift={(0,0)}]
\draw (-8,0) node[left] {$(i)$};
\Lrho(0,0,sq,w);
\draw[thick,green,rounded corners=12] (-1,0) -- (-1,1) -- (2,1);
\draw (2,1) node[right] {$z_s$};
\draw (2,0) node[right] {$s$};
\Lrho(14,0,sq,w);
\draw[thick,green,rounded corners=12] (15,0) -- (15,1) -- (16,1);
\draw (16,1) node[right] {$z_s$};
\draw (16,0) node[right] {$s$};
\draw (6,0) node[] {$\quad\quad =C_1(s,w)$};
\end{scope}
\end{tikzpicture}

\vspace*{5mm}
\begin{tikzpicture}[scale=0.4]
\begin{scope}[shift={(0,0)}]
\draw (-8,0) node[left] {$(ii)$};
\Lrho(0,0,s,w);
\draw[thick,green,rounded corners=12] (-2,1) -- (1,1) -- (1,0);
\draw (-2,1) node[left] {$z_s$};
\draw (2,0) node[right] {$sq^{-1}$};
\Lrho(14,0,s,w);
\draw[thick,green,rounded corners=12] (12,1) -- (13,1) -- (13,0);
\draw (12,1) node[left] {$z_s$};
\draw (16,0) node[right] {$sq^{-1}$};
\draw (6,0) node[] {$\quad\quad =C_2(s,w)$};
\end{scope}
\end{tikzpicture}

\vspace*{5mm}
\begin{tikzpicture}[scale=0.4]
\begin{scope}[shift={(0,0)}]
\draw (-8,0) node[left] {$(iii)$};
\Lrhobar(0,0,sq,w);
\draw[thick,green,rounded corners=12] (-1,0) -- (-1,1) -- (2,1);
\draw (2,1) node[right] {$z_s$};
\draw (2,0) node[right] {$s$};
\Lrhobar(14,0,sq,w);
\draw[thick,green,rounded corners=12] (15,0) -- (15,1) -- (16,1);
\draw (16,1) node[right] {$z_s$};
\draw (16,0) node[right] {$s$};
\draw (6,0) node[] {$\quad\quad =C_1(s,w)$};
\end{scope}
\end{tikzpicture}

\vspace*{5mm}
\begin{tikzpicture}[scale=0.4]
\begin{scope}[shift={(0,0)}]
\draw (-8,0) node[left] {$(iv)$};
\Lrhobar(0,0,s,w);
\draw[thick,green,rounded corners=12] (-2,1) -- (1,1) -- (1,0);
\draw (-2,1) node[left] {$z_s$};
\draw (2,0) node[right] {$sq^{-1}$};
\Lrhobar(14,0,s,w);
\draw[thick,green,rounded corners=12] (12,1) -- (13,1) -- (13,0);
\draw (12,1) node[left] {$z_s$};
\draw (16,0) node[right] {$sq^{-1}$};
\draw (6,0) node[] {$\quad\quad =C_2(s,w)$};
\end{scope}
\end{tikzpicture}

\vspace*{5mm}

We have one final pair of fusion relations, now concerning the composite operator $B_{r';ss'}:\Omega_{rr'} \ot \Omega_{ss'}\rightarrow \Omega_{rs} \ot \Omega_{s'r'}$ defined in Equation \ref{eq:ABdef}.
 \begin{prop}\label{prop:Vfusion}
We have the following equalities 
    \ben &&(i)\quad (B_{r';ss'}\ot \id) (\id \ot \check{\bf L}_{\Omega_{ss'}}(z_{r'}) )(\bar{I}_{r'}\ot \id)= E_1(r',s,s')\,(\id \ot \bar{I}_{r'})\, B_{r'q;ss'} 
     : W\ot W \rightarrow W\ot W\ot V\\[2mm]
    &&(ii)\quad (\id \ot\bar{T}_{r'})  (B_{r',ss'}\ot \id)(\id \ot \check{\bf L}_{\Omega_{s,s'}}(z_{r'})) = E_2(r',s,s') \,  B_{r'q^{-1},ss'}(\bar{T}_{r'}\ot \id)
    : W\ot V\ot W \rightarrow W\ot W,\een
where $\check{\bf L}_{\Omega_{ss'}}(z)$ is the $\uqt$ intertwiner $:\pi_z\ot \Omega_{ss'}\rightarrow \Omega_{ss'}\ot \pi_z$ defined above, and the coefficients are 
given by
\ben E_1(r',s,s')=\frac{\mu_s (q^2 z_{r'}^2 -z_s^2)(q^{2}x_{r'} \mu_{r'} \mu_{s'}-y_{s'})}{\mu_s x_s-\mu_{r'} x_{r'} q^{2}},\quad
 E_2(r',s,s')=\frac{\mu_{s'} q^{{2}}(z_{r'}^2-z_{s'}^2)( x_{r'} \mu_{r'}-x_s \mu_s)}{ y_{s'}-x_{r'} \mu_{r'}\mu_{s'}}
 .\een

\end{prop}

\vspace*{2mm}
\noindent The corresponding pictures for the equalities in Proposition \ref{prop:Vfusion} are
\vspace*{2mm}

\begin{tikzpicture}[scale=0.42]
\draw (-8,0) node[left] {$(i)$};
\Vmatrix(0,0);
\draw[thick,green,rounded corners=12] (-1,0) -- (-1,1) -- (2,1);
\draw (-2,-0.4) node[left] {$r$ \,} ;
\draw (-2,0.2) node[left] {$r'q$};
\draw (-0.2,2) node[above] {$s$};
\draw (0.5,2) node[above] {$s'$};
\draw (-0.5,-2) node[below] {$r$};
\draw (0.2,-2) node[below] {$s$};
\draw (2,-0.4) node[right] {$s'$};
\draw (2,0.2) node[right] {$r'$};
\draw (2,1) node[right] {$z_{r'}$};
\draw (6.5,0) node[] {$\quad\quad =E_1(r',s,s')$};
\begin{scope}[shift={(14,0)}]
\Vmatrix(0,0);
\draw[thick,green,rounded corners=12] (1,0) -- (1,1) -- (2,1);
\draw (-2,-0.4) node[left] {$r$ \,};
\draw (-2,0.2) node[left] {$r'q$};
\draw (-0.2,2) node[above] {$s$};
\draw (0.5,2) node[above] {$s'$};
\draw (-0.5,-2) node[below] {$r$};
\draw (0.2,-2) node[below] {$s$};
\draw (2,-0.4) node[right] {$s'$};
\draw (2,0.2) node[right] {$r'$};
\draw (2,1) node[right] {$z_{r'}$};
\end{scope}
\begin{scope}[shift={(0,-7)}]
\draw (-8,0) node[left] {$(ii)$};
\Vmatrix(0,0);
\draw[thick,green,rounded corners=12] (-2,1) -- (1,1) -- (1,0);
\draw (-2,-0.4) node[left] {$r\,$};
\draw (-2,0.2) node[left] {$r'$};
\draw (-0.2,2) node[above] {$s$};
\draw (0.5,2) node[above] {$s'$};
\draw (-0.5,-2) node[below] {$r$};
\draw (0.2,-2) node[below] {$s$};
\draw (2,-0.4) node[right] {$s'$};
\draw (2,0.2) node[right] {$r' q^{-1}$};
\draw (-2,1) node[left] {$z_{r'}$};
\draw (6.5,0) node[] {$\quad\quad =E_2(r',s,s')$};
\end{scope}
\begin{scope}[shift={(14,-7)}]
\Vmatrix(0,0);
\draw[thick,green,rounded corners=12] (-2,1) -- (-1,1) -- (-1,0);
\draw (-2,-0.4) node[left] {$r\,$};
\draw (-2,0.2) node[left] {$r'$};
\draw (-0.2,2) node[above] {$s$};
\draw (0.5,2) node[above] {$s'$};
\draw (-0.5,-2) node[below] {$r$};
\draw (0.2,-2) node[below] {$s$};
\draw (2,-0.4) node[right] {$s'$};
\draw (2,0.2) node[right] {$r' q^{-1}$};
\draw (-2,1) node[left] {$z_{r'}$};
\end{scope}
\end{tikzpicture}
\vspace*{2mm}

\noindent The proof of Proposition \ref{prop:Vfusion} is given in Appendix \ref{app:Vfusionproof}.

\section{Transfer Matrices and Q-operators}\label{sec:TQ}
In this section we take traces of products of the different L-operators and of $A,B$ defined in Equation (\ref{eq:ABdef}) in order to define and relate transfer matrices and Q-operators acting on quantum spaces which are either $V^{\ot M}$ or $W^{\ot M}$. We consider the two cases separately.
\subsection{Quantum Space $V^{\ot M}$ : the 6--Vertex Model}
The twisted transfer matrix of the 6-vertex model is defined as follows:
\bea T(z):=\hbox{Tr}_{V^{0}}\left( (\pi_z(t_1^\alpha)\ot \id) R^{0M}(z) \cdots R^{02}(z) R^{01}(z)\right):V^{\ot M}\rightarrow V^{\ot M}.\label{eq:T6V}\eea
Note, that in this definition we use $R=P \check{R}$ for notational simplicity. The superscripts indicate on which pair of spaces in $V^{\ot M+1}=V^{0}\ot V^{1}\ot \cdots \ot V^{M}$ the R-matrix acts. We will discuss the twist parameter $\alpha$ below. 
The graphical realization in terms of $\check{R}$ that we use is simply
\vspace*{5mm}

\begin{tikzpicture}[scale=0.3]
 \foreach \r in {0,...,5}{
\draw [thick,green](3*\r,2) -- (3*\r,-2);
\draw (3*\r,2) node[above] {$w$};
}
\draw [thick,green](-2,0) -- (17,0);
\draw (-4,0) node[left] {$T(z/w)\sim$};
\draw (-2,0) node[left] {$z$};
\end{tikzpicture}

\vspace*{5mm}
\noindent Here, and from now on, it should be understood that there is an implied trace associated with the horizontal auxiliary space in the graphical representation that we use for transfer matrices and Q-operators. For clarity, also suppress the twist operator $\pi_z(t_1^\alpha)$ that acts on the horizontal auxiliary space from the pictures.

Now we can change the auxiliary representation $\pi_z$, constructed in terms of the vector space $V$, to any of the four other representations $\rho_{r}$, $\bar{\rho}$, $\Omega_{rs}$, $\varphi_c$ associated with the vector space $W$.
Let us start with $\rho_{r}$. Then the R-matrix $R(z/w):\pi_z\ot \pi_w \rightarrow \pi_z\ot \pi_w$ appearing appearing in (\ref{eq:T6V}) is replaced by the corresponding L-operator $L_{\rho_r}(w): \rho_r\ot \pi_w\rightarrow \rho_r\ot \pi_w$.
In this way we define a new operator
\ben Q_{\rho_r}(w):=\hbox{Tr}_{W}\left( (\rho_r(t_1^\alpha) \ot \id) L_{\rho_r}^{M}(w) \cdots L_{\rho_r}^{2}(w) L_{\rho_r}^{1}(w)\right):V^{\ot M}\rightarrow V^{\ot M},\een
with $L_{{\rho}_r}^{i}$ acting non-trivially on $W\ot V^{i}$. The picture is 

\vspace*{2mm}
\begin{tikzpicture}[scale=0.3]
 \foreach \r in {0,...,5}{
\draw [thick,green](3*\r,2) -- (3*\r,-2);
\draw (3*\r,2) node[above] {$w$};
}
\draw [thick,red](-2,0) -- (17,0);
\draw (-4,0) node[left] {$Q_{\rho_r}(w)\sim$};
\draw (-2,0) node[left] {$r$};
\end{tikzpicture}
\vspace*{2mm}

Let us now discuss the reason for including the twist parameter $\alpha$ in this and other definitions. First, we recall that in the case of generic $q$, we are forced to introduced such a twist in order to regularise the infinite-dimensional trace involved in the definition of the Q-operator (see for example, \cite{Korff_2004b}). The cyclic representations in this paper are of course $N$-dimensional, and at first sight it would appear that the twist is unnecessary. However, with cyclic representations we encounter a different issue: if we make a particular choice of the representation of the matrices $X,Z\in GL(N,\mathbb{C})$ to be 
$X_{ij}=\delta_{i,\hbox{mod}(j+1,N)}$, $Z_{ij}= \delta_{i,j}q^i$, for $i,j\in\{0,1,\cdots,N-1\}$, then it follows from basic properties of primitive roots of unity that we have
\ben
\hbox{Tr}_{W}\left( X^n Z^m\right)=N \delta_{n\equiv 0}\delta_{m\equiv 0},
\een
where $\delta_{n\equiv 0}$ is defined to be equal to 1 if $n=0$ mod$(N)$, and $0$ otherwise. Requiring $n\equiv 0$ is equivalent to `charge conservation mod $N$' which is to be expected for transfer matrices associated with cyclic representations. Requiring that $m=0$ however means that the corresponding trace vanishes on all spin sectors apart from $0$ mod $N$. The twisted trace is introduced in order to stop this phenomenon.
Using the definition $\rho_r(t_1)=q\mu_r Z^2$, we see that, provided we choose $2\alpha\notin \mathbb{Z}$, we have
\ben
\hbox{Tr}_{W}\left( \rho_r(t_1^\alpha) X^n Z^m\right)=\delta_{n\equiv 0} \;q \mu_r  \sum\limits_{p=0}^{N-1} q^{(2\alpha +m) p}= \delta_{n\equiv 0}\; q \mu_r \frac{1-q^{2\alpha N}}{1-q^{2\alpha +m}}.
\een
Thus we regularize our traces by choosing the specified twist insertion in all cases, with the requirement that $2\alpha\in \mathbb{R}\backslash \mathbb{Z}$.

In a similar way we define three more operators $V^{\ot M}\rightarrow V^{\ot M}$:
\vspace*{2mm}
 
\begin{tikzpicture}[scale=0.3]
 \foreach \r in {0,...,5}{
\draw [thick,green](3*\r,2) -- (3*\r,-2);
\draw (3*\r,2) node[above] {$w$};
}
\draw [thick,blue](-2,0) -- (17,0);
\draw (-4,0) node[left] {$\sim$};
\draw (-2,0) node[left] {$r$};
\draw (-6,0) node[left] {$Q_{\bar{\rho}_r}(w):=\hbox{Tr}_{W}\left( (\bar{\rho}_r(t_1^\alpha) \ot \id)L_{\bar{\rho}_r}^{M}(w) \cdots L_{\bar{\rho}_r}^{2}(w) L_{\bar{\rho}_r}^{1}(w)\right)$};
\end{tikzpicture}

\vspace*{5mm}

\begin{tikzpicture}[scale=0.3]
 \foreach \r in {0,...,5}{
\draw [thick,green](3*\r,2) -- (3*\r,-2);
\draw (3*\r,2) node[above] {$w$};
}
\draw [thick,blue](-2,0) -- (17,0);
\draw [thick,red](-2,-0.3) -- (17,-0.3);
\draw (-4,0) node[left] {$\sim$};
\draw (-2,0.2) node[left] {$s$};
\draw (-2,-0.5) node[left] {$r$};
\draw (-5.5,0) node[left] {$T_{\Omega_{rs}}(w):=\hbox{Tr}_{W}\left( (\Omega_{rs}(t_1^\alpha) \ot \id)L_{\Omega_{rs}}^{M}(w) \cdots L_{\Omega_{rs}}^{2}(w) L_{\Omega_{rs}}^{1}(w)\right)$};
\end{tikzpicture}

\vspace*{5mm}

\begin{tikzpicture}[scale=0.3]
 \foreach \r in {0,...,5}{
\draw [thick,green](3*\r,2) -- (3*\r,-2);
\draw (3*\r,2) node[above] {$w$};
}
\draw [thick,dashed](-2,0) -- (17,0);
\draw (-4,0) node[left] {$\sim$};
\draw (-6,0) node[left] {
$T_{\varphi_c}:=\hbox{Tr}_{W}\left( (\varphi_c(t_1^\alpha) \ot \id) L_{\varphi}^{M} \cdots L_{\varphi}^{2} \,L_{\varphi}^{1}\right)$
};
\end{tikzpicture}
\vspace*{3mm}

\noindent The L-operator $L_\varphi$, given by (\ref{eq:Ldefs}), is particularly simple: it is independent of $w$, diagonal and equal to
\ben L_\varphi=\begin{pmatrix} q^{-1}Z^{-1}  & 0\\0 & Z \end{pmatrix}, \een
which immediately gives
\ben &&T_{\varphi_c}= q^{-(S_z+M)/2}\left(\frac{q}{c}\right)^\alpha 
\frac{1-q^{2\alpha N}}{1-q^{2\alpha - S_z}},
\\&&
\hbox{where}\quad S_z=\sum_{i=1}^M \sigma_z^i, \quad \hbox{with}\quad \sigma_z=\begin{pmatrix} 1&0\\0&-1\end{pmatrix}.
\een

We now present various relations between these T and Q operators that follow directly from the properties of L-operators obtained in Section \ref{sec:f&f}. First of all, we note that the operators $Q_{\rho_r}(w)$, $Q_{\bar{\rho}_r}(w)$ and $T_{\Omega_{rs}}(w)$ do not commute with the total spin operator $S_z$; it is a consequence of the relations $X^N=\id$ that these operators only conserve spin modulo $N$ as we have explained above. This feature of cyclic representations was noted in \cite{BS90}.

\begin{prop}\label{prop:QTcom}
\ben [T(z/w),Q_{\rho_{r}}(w)]= [T(z/w),Q_{\bar{\rho}_{r}}(w)]=0.\een
\end{prop}
\begin{proof}
The statement follows from Proposition \ref{prop:RLL} applied to the two auxiliary spaces in each case, along with the properties 
\ben 
&&\check{L}_{\rho_r}(z) \circ \left( \rho_r\ot \pi_z\right) \Delta(t_1)= \left(  \pi_z\ot \rho_r\right) \Delta(t_1) \circ \check{L}_{\rho_r}(z),\\ &&\check{L}_{\bar{\rho}_r}(z) \circ \left( \rho_r\ot \pi_z\right) \Delta(t_1)= \left(  \pi_z\ot \bar{\rho}_r\right) \Delta(t_1) \circ \check{L}_{\bar{\rho}_r}(z)
\een
for the twist factors, which follow from the fact that the L-operators are $\uqb$ intertwiners.
\end{proof}

\begin{prop}\label{prop:Tfact}
We have the factorization property:
\ben T_{\Omega_{rs}(w)}= Q_{\rho_r}(w) Q_{\bar{\rho}_s}(w) T_{\varphi_c}^{-1}.
\een
\end{prop}
\begin{proof}
This follows directly from Proposition \ref{prop:Lproperties} and Theorem \ref{prop:factor}.
The graphical proof (with suppression of twist insertions) is simply
\vspace*{4mm}

\begin{tikzpicture}[scale=0.3]
 \foreach \r in {0,...,5}{
\draw [thick,green](3*\r,2) -- (3*\r,-2);
}
\draw [thick,blue](-2,-0.5) -- (17,-0.5);
\draw [thick,red](-2,-0.8) -- (17,-0.8);
\draw [dashed](-2,0.8) -- (17,0.8);
\draw (-6,0) node[left] {$\sim$};
\draw (-8,0) node[left] {$T_{\Omega_{rs}}(w) T_{\varphi_c} $};
\begin{scope}[shift={(0,-5)}]
\foreach \r in {0,...,5}{
\draw [thick,green](3*\r,2) -- (3*\r,-2);
}
\draw [thick,blue](-2,-0.5) -- (17,-0.5);
\draw [thick,red](-2,-0.8) -- (17,-0.8);
\draw [dashed](-2,0.8) -- (17,0.8);
\draw (-6,0) node[left] {$=$};
\Oiso(17,0);
\Oisoinv(19.1,0);
\end{scope}
\begin{scope}[shift={(0,-10)}]
\foreach \r in {0,...,5}{
\draw [thick,green](3*\r,2) -- (3*\r,-2);
}
\draw [thick,red](-2,-0.8) -- (17,-0.8);
\draw [thick,blue](-2,0.8) -- (17,0.8);
\draw (-6,0) node[left] {$=$};
\Oiso(-2,0);
\Oisoinv(17,0);
\end{scope}
\begin{scope}[shift={(0,-15)}]
\foreach \r in {0,...,5}{
\draw [thick,green](3*\r,2) -- (3*\r,-2);
}
\draw [thick,red](-2,-0.8) -- (17,-0.8);
\draw [thick,blue](-2,0.8) -- (17,0.8);
\draw (-6,0) node[left] {$=$};
\draw (19,0) node[right] {$\sim\quad Q_{\rho_r}(w) Q_{\bar{\rho}_s}(w) $};
\end{scope}
\end{tikzpicture}
\vspace*{5mm}

\noindent where Proposition \ref{prop:Lproperties} is used between the second and third diagrams, and cyclicity of the trace is used in the final step.
\end{proof}

\begin{prop}\label{prop:TQV}
\ben
&(i)\quad Q_{{\rho}_s}(w) T(z_s/w) = C_1^M(s,w) Q_{{\rho}_{sq}}(w) + C_2^M(s,w) Q_{{\rho}_{sq^{-1}  }}(w),\\[2mm]
&(ii)\quad 
Q_{\bar{\rho}_s}(w) T(z_s/w) = C_1^M(s,w) Q_{\bar{\rho}_{sq}}(w) + C_2^M(s,w) Q_{\bar{\rho}_{sq^{-1}  }}(w).\\[2mm]
\een
\end{prop}
\begin{proof}The proof follows from the SESs of Proposition \ref{prop:SESs} $(i)$ and $(ii)$ and Proposition \ref{prop:Lfusion}. We also make use of the standard linear algebra result that given a short exact sequence $0\rightarrow A \rightarrow B \rightarrow C\rightarrow 0$, the trace over $B$ of an operator acting on $B$ can we written as the sum of traces of corresponding operators over $A$ and $C$. Details of this result as applied in this context can be found in \cite{VW20}.
\end{proof}

Relations $(i)$ and $(ii)$ of Proposition \ref{prop:TQV} both contain the same factors $C_1^M(s,w)=(b(z_s/w)/q)^M$ and 
$C_2^M(s,w) =(q a(z_s/w))^M$. These are precisely the coefficients those that occur in the standard TQ relations of the 6-vertex model - which are in turn used to obtain the Bethe equations (see for example \cite{JMS21}). The apparent difference with the standard TQ relations is the more complicated 
dependence of the Q operators in $(i)$ and $(ii)$ on $s\in\cC_k$ and $w$. However,  
the TQ relation may be brought into the standard form by making use of the following `gauge transformation' of the matrices $U_s(w)$ and $V_s(w)$ defined in Equation (\ref{eq:UVdef}):
\ben 
U_s(w) &=& w\, \aleph_s U(z_s/w,\mu_s) \aleph_s^{-1},\;\;\hbox{where}\;\; U(z,\mu):=\begin{pmatrix} 1& z\\z&1\end{pmatrix} 
\begin{pmatrix} 1& 0\\0&\mu\end{pmatrix} \;\; \hbox{and}\;\;
\aleph_s:=\begin{pmatrix} 1&0\\0 & \sqrt{\frac{\kappa_1 y_s}{\kappa_0 x_s}}\end{pmatrix}, 
\\
V_s(w) &=& w\,\beth_s \, V(z_s/w,\mu_s) \beth_s^{-1},\;\;\hbox{where}\;\;
V(z,\mu):=\begin{pmatrix} 1& 0\\0&\mu\end{pmatrix} \begin{pmatrix} -q& z\\z&-1\end{pmatrix}\;\;\hbox{and}\;\;
\beth_s:=\begin{pmatrix} 1&0\\0 & \sqrt{\frac{\kappa_1 x_s}{\kappa_0 y_s}}\end{pmatrix}.\een
Let us define $${L}(z,\mu)=\{U(z,\mu),\id\},\quad {\widebar{L}}(z,\mu)=\{V(z,\mu),\id\},
$$
using the bracket notation of Equation (\ref{eq:ABnotn}). Then it follows that we have
\bea
{L}_{\rho_s}(w)=w\, \aleph_s  {L}(z_s/w,\mu_s) \aleph_s^{-1},\quad
{L}_{\bar{\rho}_s}(w)=w\, \beth_s  {\widebar{L}}(z_s/w,\mu_s) \beth_s^{-1}.\label{eq:gtransL}\eea
Using the transformed L-operators $L(z,\mu)$ and $\bar{L}(z,\mu)$, we can define modified Q operators by
\ben Q(z,\mu_s):&=&\Tr_W(  (\rho_s(t_1^\alpha) \ot \id)   {L}^M(z,\mu_s) \cdots {L}^2(z,\mu_s) {L}^1(z,\mu_s) ),\\
\widebar{Q}(z,\mu_s):&=&\Tr_W((\bar{\rho}_s(t_1^\alpha)\ot \id)  \widebar{L}^M(z,\mu_s) \cdots \widebar{L}^2(z,\mu_s) \widebar{L}^1(z,\mu_s) ),
\een
It is clear, from the form of $U(z,\mu)$ and $V(z,\mu)$, that $Q(z,\mu)$ and $\widebar{Q}(z,\mu)$ are polynomial in $z$.
It then follows from (\ref{eq:gtransL}) that we have
\bea Q(z_s/w,\mu_s)&=&w^{-M} \left(\sqrt{\frac{\kappa_1 y_s}{\kappa_0 x_s}}\right)^{\frac{S_z}{2}} Q_{{\rho}_s}(w) \left(\sqrt{\frac{\kappa_1 y_s}{\kappa_0 x_s}}\right)^{-\frac{S_z}{2}},\\
\widebar{Q}(z_s/w,\mu_s)&=&w^{-M} \left(\sqrt{\frac{\kappa_1 x_s}{\kappa_0 y_s}}\right)^{\frac{S_z}{2}} Q_{\bar{\rho}_s}(w) \left(\sqrt{\frac{\kappa_1 x_s}{\kappa_0 y_s}}\right)^{-\frac{S_z}{2}}\label{eq:Qsimp}.
\eea
Hence it then follows from Proposition \ref{prop:TQV} and the property $[T(z),S_z]=0$ that $Q(z,\mu)$ and $\widebar{Q}(z,\mu)$ satisfy
the standard TQ relations\footnote{The dependence of the $Q$ operator on the parameter $\mu$, the algebraic necessity of this parameter, and an explanation how the dependence may be removed is explained in \cite{VW20}}
\bea
Q(z,\mu) T(z) &=& (b(z)/q)^L \,Q(qz,q\mu) + (q a(z))^L \,Q(q^{-1} z,q^{-1}\mu),\nonumber\\
\widebar{Q}(z,\mu) T(z) &=& (b(z)/q)^L \,\widebar{Q}(qz,q\mu) + (q a(z))^L \,\widebar{Q}(q^{-1} z,q^{-1}\mu).
\label{eq:TQstandard}\eea

Before completing this section, we note that $T_{\Omega_{rs}}$ also satisfies the relation
\bea T_{\Omega_{r,s}}(w) T(z_s/w) = C_1^M(s,w) T_{\Omega_{r,sq}}(w) + C_2^M(s,w) T_{\Omega_{r,sq^{-1}  }}(w).\label{eq:TOmegaQ}\eea
This follows either by the observation that it addition to Proposition \ref{prop:Lfusion}, we also have that
\ben  && (\check{L}_{\Omega{rs}}(w)\ot \id)(\id\ot \check{R}(z_s/w))(\bar{I}_{s}\ot\id )= C_1(s,w) (\id \ot \bar{I}_{s}) \check{L}_{\Omega_{r,sq}}(w)= : W\ot V \rightarrow V\ot W\ot V,\\
 &&(\id \ot\bar{T}_s)  (\check{L}_{\Omega_{r,s}}(w)\ot \id)(\id \ot \check{R}(z_s/w)) = C_2(s,w) \, \check{L}_{\Omega_{r,sq^{-1}}}(w) 
 (\bar{T}_s\ot \id): W\ot V \ot V \rightarrow V\ot W ,
\een
(which can be proved either by appealing to uniqueness of the corresponding $\uqb$ intertwiner or just by checking the relation), or directly from Propositions \ref{prop:Tfact} and \ref{prop:TQV} $(ii)$ plus the observation that $T_{\varphi_c}$ commutes with $Q_{\rho_r}$ and $Q_{\bar{\rho}_s}$.
Thus we have that $T_{\Omega_{rs}}(w)$ satisfies the same relations (\ref{eq:TOmegaQ}) as the Q operator of the 6-vertex model. This fact was observed in \cite{BS90}.  One advantage of considering  $T_{\Omega_{rs}}(w)$  is that one can establish its invariance under an infinite-dimensional group action known as the quantum coadjoint action. This was the approach taken in \cite{Korff_2003a}.

\subsection{Quantum Space $W^{\ot M}$ : the $\tau_2$ and Chiral Potts Models}\label{sec:Wquantum}
Now let us define an operator $\cT(z):W^{\ot M}\rightarrow W^{\ot M}$ as follows:\vspace*{5mm}

\begin{center}
\begin{tikzpicture}[scale=0.3]
\foreach \r in {0,...,5}{
\LOmegainv(3*\r,0,,s,s');
}
\draw(-2,0) node[left] {$z$};
\draw(-36,0) node[right] {$\cT(z)=\hbox{Tr}\left((\pi_z(t_1^\alpha)\ot \id)\, 
{\bf L}_{\Omega_{ss'}}^M(z) \cdots {\bf L}_{\Omega_{ss'}}^2(z) {\bf L}^1_{\Omega_{ss'}}(z)
\right)\quad\sim$};
\end{tikzpicture}
\end{center}
\vspace*{5mm}
where ${\bf L}_{\Omega_{rs}}(z)=P \check{\bf L}_{\Omega_{rs}}(z)$ is the operator $V\ot W\rightarrow V\ot W$ defined by \ref{eq:Lbfdef} and given explicitly by Equation (\ref{eq:Lbfmatrix}); the superscript $i$ as always indicates that the operator acts on the $i$'th space $W$ in the tensor product $W^{\ot M}$. The operator $\cT(z)$ is the transfer matrix of the $\tau_2$ model as defined in \cite{Baxter:1989vv,BPY88,BS90}.

Now we use the operators $B_{r';ss'}=P \check{B}_{r';ss'}$ and $A_{rs;s'}=P \check{A}_{rs;s'}$ specified by Equation \ref{eq:ABdef} to define associated operators on $W^{\ot M+1}$ by
\ben {\mathcal B}_{r';ss'}&=&
B_{r';ss'}^{0M} \cdots B_{r';ss'}^{02} B_{r';ss'}^{01}\\
{\mathcal A}_{rs;s'}&=&
A_{rs;s'}^{0M}\cdots A_{rs;s'}^{02} A_{rs;s'}^{01}.\een
Then, we define the following operator $\cQ_{r';ss'}:W^{\ot M}\rightarrow W^{\ot M}$:
\bea
\begin{tikzpicture}[scale=0.3]
\draw (-2,-0.6) node[left] {$r$};
\draw (-2,0.3) node[left] {$r'$};
\foreach \r in {0,...,5}{
\Vmatrix(3*\r,0); 
\draw (-0.3+3*\r,2) node[above] {$s$};
\draw (0.55+3*\r,2) node[above] {$s'$};
}
\foreach \r in {1,...,5}{
\draw (-0.55+3*\r,-1.8) node[below] {$s'$};
\draw (0.25+3*\r,-2.2) node[below] {$s$};
}
\draw (17,-0.5) node[right] {$s'$};
\draw (17,0.2) node[right] {$r'$};
\draw (-0.55,-2.2) node[below] {$r$};
\draw (0.25,-2.2) node[below] {$s$};
\draw(-27,0) node[right] {$\cQ_{r';ss'}=\hbox{Tr}_{W^0}\left((\Omega_{rr'}(t_1^\alpha) \ot \id)
{\mathcal B}_{r';ss'}
\right)\sim$};
\end{tikzpicture}
\label{eq:QB}\eea

\begin{prop}\label{prop:TQW}
\ben
{\cQ}_{r';ss'} \cT(z_{r'}) = E_1^M(r';ss') \cQ_{r'q;ss'} + 
E_2^M(r';ss') \cQ_{r'q^{-1};ss'}
.\een
\end{prop}
\begin{proof} The proof follows from the SES of Proposition \ref{prop:SESs} $(iii)$ and from Proposition \ref{prop:Vfusion}.
\end{proof}

\noindent In this way, we see that $\cQ_{r';ss}$ acts as a Q-operator for the $\tau_2$ model.  
The connection of this Q-operator with the chiral Potts model is as follows:
the transfer matrix of the chiral Potts model is by definition given by
\ben \cT^{CP}=\Tr_{W^0}( (\Omega_{rr'}(t_1^\alpha) \ot \id) R(rr';ss')^{0M} \cdots R(rr';ss')^{02} R(rr';ss')^{01} ).\een
It follows from the factorization formula (\ref{eq:Rfact}) that we have 
\bea \cT^{CP}=\Tr_{W^0}( (\Omega_{rr'}(t_1^\alpha) \ot \id)\, \cA_{rs;s'}\circ P^{(M)} \circ \cB_{r';ss'} ),\label{eq:TCPAB}\eea
where $P^{(M)}:W^{\ot M+1}\rightarrow W^{\ot M+1}$ is defined by $P^{(M)}(a\ot b_1\ot b_2\cdots b_M)= (b_1\ot b_2\cdots b_M\ot a)$.
The corresponding picture is

\vspace*{5mm}

\begin{tikzpicture}[scale=0.3]
\draw (-2,-0.6) node[left] {$r\,$};
\draw (-2,0.3) node[left] {$r'$};
\draw (20,-0.5-3) node[right] {$r$};
\draw (20,0.4-3) node[right] {$r'$};
 \foreach \r in {0,...,5}{
\Vmatrix(3*\r,0);
\Umatrix(3*\r+2.7,-3);
}
\lcorner(0,-2.7);
\rcorner(17.7,-0.3);
\foreach \r in {0,...,5}{
\draw (-0.3+3*\r,2) node[above] {$s$};
\draw (0.55+3*\r,2) node[above] {$s'$};
\draw (-0.55+3+3*\r,-2.2-3) node[below] {$s$};
\draw (0.25+3+3*\r,-1.8-3) node[below] {$s'$};
}
\draw(-10,0) node[right] {$\cT^{CP}\sim$};

\end{tikzpicture}

It was shown in \cite{BS90} that there was a relation between the Q-operator of the $\tau_2$ model (found in \cite{BS90} by explicitly solving of TQ relations) and the half-monodromy matrix of the chiral Potts model. This connection was studied in further detail in \cite{Roan_2007,Roan_2009}. The algebraic origin of this connection is expressed in the current paper by three results: the definition $\cQ_{r';ss'}=\hbox{Tr}_W\left((\Omega_{rr'}(t_1^\alpha) \ot \id)
{\mathcal B}_{r';ss'}\right)$; Proposition \ref{prop:TQW}; and the identification (\ref{eq:TCPAB}).

\section{Discussion}\label{sec:discussion}

In this section we summarize the key results of the paper and comment on their potential use and development. Our first main result is the definition of the two $\uqb$ representations $\rho_{r}$ and $\bar{\rho}_r$ and the factorization Theorem \ref{prop:factor}. This result is interesting for various reasons: it can be viewed as the analog of the generic-$q$ result (i) mentioned in Section \ref{sec:intro}; it leads directly to the L-operator factorization of Proposition \ref{prop:Lproperties} and the transfer matrix factorization of Proposition \ref{prop:Tfact}; and finally,  it allows us, via Proposition \ref{prop:STalt}, to characterize chiral Potts Boltzmann weights in
an alternative way to Proposition \ref{Rfact}. 

The next key result is the SES/fusion result of Theorem \ref{prop:SESs}. This property is manifested in terms of L-operators in the way expressed in Propositions \ref{prop:Lfusion} and \ref{prop:Vfusion}. These L-operator relations in turn lead to the TQ relations of Propositions \ref{prop:TQV} and \ref{prop:TQW} for operators acting on the quantum spaces $V^{\ot M}$ and $W^{\ot M}$. More precisely, when the quantum space is $V^{\ot M}$ we show that choosing the auxiliary space to be either $\rho$ or $\bar{\rho}$ leads to operators $Q(z,\mu)$ and $\widebar{Q}(z,\mu)$ that satisfy the expected TQ relations (\ref{eq:TQstandard}) of the 6-vertex model . These operators commute with the 6-vertex model transfer matrix and are polynomial in $z$. Thus they provide a construction of the Q-operators in the $q^N=1$ case of the 6-vertex model. This is simpler than the construction in the generic-$q$ case that requires an infinite-dimensional auxiliary space as well as regularization of the trace by the introduction of a Cartan element. We also show in Proposition \ref{prop:Tfact} that the transfer matrix $T_{\Omega_{rs}}(z)$ factorizes as a product of our two Q-operator, and view this as an analogue of the factorization formula (\ref{eq:introfact}) of the Verma module transfer matrix in the case of $q$ generic \cite{BLZ97}.

For the case when the quantum space is $W^{\ot M}$, we construct an operator $\cQ_{r';ss'}$ by choosing the auxiliary space as $\Omega_{rr'}$. 
This operator satisfies TQ relations with the transfer matrix $\cT(z)$ of the $\tau_2$ model, as expressed in Proposition \ref{prop:TQW}. In this case, the shifts in the Q-operator are expressed as $r'\rightarrow r'q^{\pm 1}$ which are automorphisms of the curve $\cC_k$. Our $\cQ_{r';ss'}$ is constructed as a trace over the operator $\cB_{r';ss'}$ which is also a building block (in fact one half of the monodromy matrix)  of the transfer matrix $\cT^{CP}$ of the chiral Potts model as expressed in Equation \ref{eq:TCPAB}.

One chapter of the Q-operator story is missing so far in our discussion of the $q^N$ case: a feature of the generic-$q$ case is that the infinite-dimensional Verma module module $\nu^\mu_z$ mentioned in Section \ref{sec:intro} has an infinite-dimensional submodule when $\mu$ takes an integer value $n$. The quotient is then isomorphic to the $n+1$ dimensional evaluation module (see for example \cite{VW20} for details). This means that the factorization formula (\ref{eq:introfact}) can be used to express the transfer matrix $T^{(n)}(z)$ associated with the $n+1$ dimensional auxiliary representation in the form 
\bea T^{(n)}(z)= \hash \cQ(zq^{-n/2})\bcQ(zq^{n/2}) - \hash \cQ(zq^{n/2})\bcQ(zq^{-n/2}).\label{eq:QQ}\eea
This reducibility does not occur for the general semi-cyclic representations of this paper. In \cite{MLP21} however, L-operators are considered for semi-cyclic representations, and used to define Q-operators that satisfy a relation of the form (\ref{eq:QQ}). The semi-cyclic limit of our representation $\Omega_{rs}$ was first studied in the paper \cite{IU92}, and it would be interesting to attempt to recover the results of \cite{MLP21} from the current paper in this limit.

Both the cyclic representations $\Omega_{rs}$ and the chiral Potts have been generalized to higher-rank cases \cite{DJMM91a,BKMS90}. Also, the generic-$q$ factorization result (i) in Section 1 has been generalized to all untwisted quantum affine algebras in \cite{HJ12,FH15}. It should be possible and would certainly be interesting  to generalize our $q^N$ factorization Theorem \ref{prop:factor} to higher rank.

As mentioned in Section \ref{sec:intro}, another potential use of our results is in the analysis of open systems. Q-operators for some generic-$q$ open systems have been constructed in terms of infinite-dimensional $\uqb$ representations in 
\cite{FSz15,BTs18,VW20,Tsuboi2021,CVW24}. The reason this construction is possible is that the recently formulated universal K-matrix for quantum affine algebras \cite{AV22a,AV22b,AV24} is itself intimately related to the Borel algebras, as well as to coideal subalgebras of the underlying quantum affine algebra. One barrier to developing this approach and constructing explicit Q-operators for open systems with arbitrary boundary conditions is the complexity of solving infinite-dimensional reflection equations. It is hoped that the finite-dimensional cyclic $\uqb$ representations introduced in this paper will be of use in this context.


\subsection*{Acknowledgments}
The author thanks Alec Cooper, Bart Vlaar, Jules Lamers, Yuan Miao and Christian Korff for useful discussions. He would also like to thank MATRIX and the organizers of the programme `Mathematics and Physics of Integrability (MPI2024)' during which part of this work was completed. Finally, he acknowledges funding from the EPSRC via grant EP/V008129/1.

\begin{appendix}
\section{Proofs from Section \ref{sec:f&f}}\label{app:proofs}
\subsection{Proof of Lemma \ref{lem:Oinverse}}\label{app:Oinvproof}
Using the fact that $q^N=1$, $\chi^N=\id$, and supposing that $\cO^{-1}(\chi) =\sum\limits_{n=0}^{N-1} b_n \chi^N$, we can then write
\ben \cO(\chi) \cO^{-1}(\chi) = (\id, \chi^{N-1},\chi^{N-2},\cdots,\chi) \, A\, 
\begin{pmatrix} b_0\\ b_{N-1}\\b_{N-2}\\ \vdots \\ b_1 \end{pmatrix},
\quad \hbox{where}\; A=\begin{pmatrix} 
a_0& a_1 & a_2& \cdots &a_{N-1}\\
a_{N-1}& a_0 & a_1& \cdots &a_{N-2}\\
a_{N-2}& a_{N-1} & a_0 &\cdots &a_{N-3}\\
\vdots &\vdots & \vdots&&\vdots\\
a_1& a_2 & a_3& \cdots &a_{0}\end{pmatrix}.\een
The requirement that $\cO(\chi) \cO^{-1}(\chi)=\id$ is then equivalent to the matrix relation
\bea A \begin{pmatrix} b_0\\ b_{N-1}\\b_{N-2}\\ \vdots \\ b_1 \end{pmatrix} =
\begin{pmatrix} 1\\ 0\\ 0\\ \vdots \\ 0 \end{pmatrix}.\label{eq:Abreln}\eea
This type of matrix $A$ consisting of cyclically permuted column vectors is know as a {\it circulant matrix} and has many nice properties (see for example \cite{Davies94}). In particular, its eigenvalues and eigenevectors are known and are respectively
\ben \lambda_j = \sum\limits_{n=0}^{N-1} a_n q^{jn}=\cO(q^j),\quad v_j=\begin{pmatrix} 1 \\ q^j \\ q^{2 j} \\ \vdots\\ q^{(N-1)j}\end{pmatrix}, \quad j\in\{0,1,\cdots,N-1\}.
\een
Hence the matrix $A$ is invertible iff $\cO(q^j)\neq 0$ for all $j\in\{0,1,\cdots,N-1\}.$ 
We find using the Landsberg-Schaar identity that
\ben \cO(1)=\sum_{n=0}^{N-1}q^{-n^2}= 
\frac{1+i^N}{1+i}\sqrt{N} .\een
The values of $\cO(q^j)$ for all $j$ then follow from the required relation $\cO(qz)=z \cO(q^{-1} z)$ for $z^N=1$. In this way we see that we indeed have $O(q^j)\neq 0$ for all $j\in\{0,1,\cdots,N-1\}$ and hence $A$ is invertible.
It follows from \ref{eq:Abreln} that the vector $\underline{b}$ is given by the first column of the matrix $A^{-1}$. 
\hfill$\Box$
\subsection{Proof of Proposition \ref{prop:Lfusion}}\label{app:Lfusionproof}
There are two approaches to proving (i)-(iv): we can either appeal to uniqueness up to a multiplicative constant of the associated $\uqb$ intertwiner, and thus identify the the two sides by checking a single matrix element; or, we can simply check the full relation as a linear algebra identity. We have done both. As an example, let us show the latter pedestrian approach in detail for the case (iii). 
\vspace*{3mm}

\noindent Statement (iii): We can express either side of the identity as a $4\times 2$ matrix with respect to the space $V$. The left-hand side minus the right-hand side is then 
\vspace*{3mm}
\ben
\left(
\begin{array}{cc}
 \frac{\kappa_0 q y_s (-a w+C_1 w+ c z_s)}{z_s Z} & \frac{\kappa_0^2 q y_s^2 (b-C_1 q)
   {X}^{-1} Z}{z_s} \\
 \frac{w (C_1 q-b) X Z^{-3}}{q} & -\frac{\kappa_0 q y_s \left(-a z_s+C_1 q^2 z+c w\right)}{z_s Z} \\
 -\frac{\mu  q X Z^{-1} \left(\kappa_0 \kappa_1 x_s y_s \left(C_1 q^2-a\right)+c w z_s\right)}{z_s} &
   \frac{\kappa_0\mu_s  q w y_s Z (C_1 q-b)}{z_s} \\
 \frac{\kappa_1\mu_s x_s (b-C_1 q) X^2 Z^{-3}}{q} & \frac{\mu_s  q X {Z}^{-1} (w z_s
   (C_1-a)+ c \kappa_0\kappa_1 x_s y_s)}{z_s} \\
\end{array}
\right)
\een
\vspace*{3mm}
where we have suppressed the arguments of the entries $(a,b,c)$ of the R-matrix $\check{R}(z_s/w)$ given by (\ref{eq:Rmatrix}), as well as those of $C_1(s,w)$. Recalling  that $z_s^2=\kappa_0\kappa_1 x_sy_s$, we see that each of the terms in the matrix is zero with the choice $C_1(s,w)=q^{-1} b(z_s/w)$.
\vspace*{2mm}

\noindent The proof of Statement (i) is very similar. The proofs of Statements (ii) and (iv) involving showing that all the entries of a $2\times 4$ are zero with the specified choice of $C_2(s,w)=q a(z_s/w)$.

\subsection{Proof of Proposition \ref{prop:Vfusion}}\label{app:Vfusionproof}
Let us define a matrix
\ben \cM(z)=\begin{pmatrix}
y_s y_{s'} \kappa_0 \kappa_1 Z^{-1} - q^2 z^2 \mu_s\mu_{s'} Z
& q z_s \kappa_0(-y_{s'} Z^{-1} + x_s \mu_s \mu_{s'}Z \\
q z \kappa_1(y_s Z^{-1}-q^2 x_{s'} \mu_s\mu_{s'} Z)& q^2 (-z^2 Z^{-1} + x_{s} x_{s'} \kappa_0\kappa_1 \mu_{s}\mu_{s'} Z
\end{pmatrix}
\een
Then the explicit expression for the inverse matrix $\check{\bf L}_{\Omega_{ss'}}(z):V\ot W\rightarrow W\ot V$ matrix introduced in Section 3 and involved in Proposition \ref{prop:Vfusion} is
\bea \check{\bf L}_{\Omega_{ss'}}(z) = \begin{pmatrix} 
\cM(z)_{00} & X^{-1} \cM(z)_{01}\\
X \cM(z)_{10} & \cM(z)_{11}.\end{pmatrix}\label{eq:Lbfmatrix}\eea
Each component $\cM(z)_{ij}$ then acts $W\rightarrow W$.

Statement (i) in the proposition involves two components with respect to the $V$ space. 
These components are equivalent to the following identities $\End(W\ot W)$:
\bea d_{r} B_{r;ss'}(\id \ot \cM(z_{r})_{00}) + B_{r;ss'} \chi^{-1} 
(\id\ot \cM(z_{r})_{01})&=&d_{r} E_1(r,s,s')  B_{rq;ss'},\label{eq:step1}\\
B_{r;ss'} 
(\id\ot \cM(z_{r})_{11})+d_{r} B_{r;ss'}\chi^{-1}(\id \ot \cM(z_{r})_{10})  &=& E_1(r,s,s') \chi^{-1} B_{rq;ss'}.\label{eq:step1-2}
\eea
with
$$ B_{r;ss'}= \sum\limits_{n,m=0}^{N-1} \overline{W}_{rs'}(n)\widehat{W}_{rs}(m)(\id\ot Z^{2n}) \chi^{m}.$$ 
Let us prove (\ref{eq:step1}). 
The simplest way to proceed is to choose a faithful representation of the 
$X,Z$ algebra on the N-dimensional space $W$. Let us choose $X,Z\in GL_N(\C)$ with $X_{ij}=\delta_{i,\hbox{mod}(j+1,N)}$, $Z_{ij}= \delta_{i,j}q^i$, for $i,j\in\{0,1,\cdots,N-1\}$. Then each component $\cM(z)_{ij}$ is a diagonal matrix with diagonal entries that we denote $\cM(z;\ell)_{ij}$ for $\ell\in\{0,1,\cdots,N-1\}$. 

The presumed identity (\ref{eq:step1}) then becomes the requirement that
\bea &&\widecheck{W}_{rs'}(\ell+m) \left(d_r  \widehat{W}_{rs}(m) \cM(z_r;\ell)_{00} + 
 \widehat{W}_{rs}(m+1) \cM(z_r;\ell)_{01}\right) \nonumber\\ &&=
\widecheck{W}_{rq,s'}(\ell+m) \widehat{W}_{rq,s}(m)\, d_{r} E_1(r,s,s') \label{eq:WWreln}\eea
for all $m,\ell\in\{0,\cdots,N-1\}$. Here $\widecheck{W}_{rs}$ is the discrete Fourier transform of $\overline{W}_{rs}$ defined in Equation (\ref{eq:WFT}.
We can re-express the left-hand-side of (\ref{eq:WWreln}) using the recursion relation (\ref{eq:WRR} as 
\bea &&\frac{\widecheck{W}_{rs'}(\ell+m) \widehat{W}_{rs}(m)}{\mu_s x_s-\mu_r x_r q^{2(m+1)}} \left[d_r  (\mu_s x_s-\mu_r x_r q^{2(m+1)} ) \cM(z_r;\ell)_{00} + (\mu_s y_r -\mu_r y_s q^{2m})
 \cM(z_r;\ell)_{01}\right] \label{eq:WWreln2}.\eea
Inserting the explicit expressions for $\cM(z;\ell)_{ij}$ the square bracket simplifies to
\bea 
[\cdots]=d_r \kappa_0 \kappa_1 \mu_s q^{-\ell} (q^2 x_r y_r -x_s y_s)(-y_{s'} + q^{2(1+\ell+m)}x_r \mu_r \mu_{s'}).
\eea
Hence our proposition is equivalent to the statement
\bea \frac{\widecheck{W}_{rq,s'}(\ell+m) \widehat{W}_{rq,s}(m)}{\widecheck{W}_{r,s'}(\ell+m) \widehat{W}_{r,s}(m)}  = \frac{\kappa_0 \kappa_1 \mu_s q^{-\ell}}{E_1(r,s,s')} \frac{(q^2 x_r y_r -x_s y_s)(-y_{s'} + q^{2(1+\ell+m)}x_r \mu_r \mu_{s'})}{\mu_s x_s-\mu_r x_r q^{2(m+1)}}\label{eq:step4}\eea
for all $m,\ell\in\{0,\cdots,N-1\}$.

To prove (\ref{eq:step4}) we use induction in $\ell$ and $m$. Firstly, we observe that, with the specified normalizations
$\widecheck{W}_{rs}(0)=\widehat{W}_{rs}(0)=1$, the statement is true for $m=\ell=0$ when we choose 
\ben E_1(r,s,s') = \frac{\mu_s (q^2 z_r^2 -z_s^2)(-y_{s'} + q^{2}x_r \mu_r \mu_{s'})}{\mu_s x_s-\mu_r x_r q^{2}}.\een
Now assume that (\ref{eq:step4}) is true for a given $(\ell,m)$.
Then using (\ref{eq:WFTRR2} we find
\ben \frac{\widecheck{W}_{rq,s'}(\ell+1+m) \widehat{W}_{rq,s}(m)}{\widecheck{W}_{r,s'}(\ell+1+m) \widehat{W}_{r,s}(m)}&=& \frac{\widecheck{W}_{rq,s'}(\ell+m) \widehat{W}_{rq,s}(m)}{\widecheck{W}_{r,s'}(\ell+m) \widehat{W}_{r,s}(m)}
\frac{y_{s'} - x_r \mu_r \mu_{s'} q^{2(\ell+2+m)}}
{q(y_{s'}-x_r \mu_r \mu_{s'} q^{2(\ell+1+m)})} \\&&=
\frac{\kappa_0 \kappa_1 \mu_s q^{-(\ell+1)}}{E_1(r,s,s')} \frac{(q^2 x_r y_r -x_s y_s)(-y_{s'} + q^{2(2+\ell+m)}x_r \mu_r \mu_{s'})}{\mu_s x_s-\mu_r x_r q^{2(m+1)}},
\een
and thus (\ref{eq:step4}) is true for $(\ell+1,m)$. 
Similarly, we can use \ref{eq:WRR} to obtain
\ben 
\frac{\widecheck{W}_{rq,s'}(\ell+m+1) \widehat{W}_{rq,s}(m+1)}{\widecheck{W}_{r,s'}(\ell+m+1) \widehat{W}_{r,s}(m+1)}&=&
\frac{\widecheck{W}_{rq,s'}(\ell+m+1) \widehat{W}_{rq,s}(m)}{\widecheck{W}_{r,s'}(\ell+m+1) \widehat{W}_{r,s}(m)}  \frac{q(\mu_s x_s - \mu_r x_r q^{2(m+1)})}{\mu_s x_s - \mu_r x_r q^{2(m+2)}} \\
&&=
\frac{\kappa_0 \kappa_1 \mu_s q^{-\ell}}{E_1(r,s,s')} \frac{(q^2 x_r y_r -x_s y_s)(-y_{s'} + q^{2(2+\ell+m)}x_r \mu_r \mu_{s'})}{\mu_s x_s-\mu_r x_r q^{2(m+2)}}.\een
Hence the statement (\ref{eq:step4}) is true for $(\ell,m+1)$. This completes of \ref{eq:step1}, and the proof of (\ref{eq:step1-2}) is very similar.
\vspace*{2mm}

Statement (ii) of Proposition \ref{prop:Vfusion} is equivalent to the two identities. 
\bea && -\chi B_{r,ss'} (\id \ot \cM(z_r)_{00}) + d_r B_{r,ss'} \chi(\id \ot \cM(z_r)_{10})=-E_2(r,s,s')B_{rq^{-1},ss'}\label{eq:id7}\\
&& -\chi B_{r,ss'} \chi^{-1} (\id \ot \cM(z_r)_{01}) + d_r B_{r,ss'}(\id \ot \cM(z_r)_{11})=d_r E_2(r,s,s') B_{rq^{-1},ss'}.\label{eq:id8}\eea
These can both be proved in a similar way to above. Let us outline the proof of (\ref{eq:id8}).
The equation is equivalent to the identity
\bea &&\left( - \widecheck{W}_{r,s'}(m+\ell-1)  \cM(z_r;\ell)_{01}+ d_r \widecheck{W}_{r,s'}(m+\ell)\cM(z_r;\ell)_{11}\right) \widehat{W}_{rs}(m) \nonumber\\&& = d_r E_2(r,s,s') \widecheck{W}_{rq^{-1},s'}(m+\ell) \widehat{W}_{rq^{-1},s}(m).\label{eq:id9}
\eea
After using the recursion relation (\ref{eq:WFTRR2}), the left-hand-side of (\ref{eq:id9}) simplifies to 
\ben \left[  - (y_r-x_{s'} \mu_r \mu_{s'} q^{2(m+\ell)}) \cM(z_r;\ell)_{01}  + d_{r} (y_{s'}-x_r \mu_r\mu_{s'} q^{2(m+\ell)} )   \cM(z_r;\ell)_{11}\right]  \frac{\widecheck{W}_{r;s'}(m+\ell)\widehat{W}_{rs}(m)}{y_{s'}-x_r \mu_r\mu_{s'} q^{2(m+\ell)}}  
\een
The square bracket now simplifies to
\ben \left[\cdots\right] = d_r \mu_{s'} \kappa_0 \kappa_1 q^{{2+\ell}}(x_r y_r-x_{s'} y_{s'})(q^{2m} x_{r} \mu_r-x_s \mu_s),\een
and Equation (\ref{eq:id8}) is equivalent to 
\ben \frac{\widecheck{W}_{rq^{-1},s'}(m+\ell) \widehat{W}_{rq^{-1},s}(m)}{\widecheck{W}_{r;s'}(m+\ell)\widehat{W}_{rs}(m)}= \frac{\mu_{s'} \kappa_0 \kappa_1 q^{{2+\ell}}(x_r y_r-x_{s'} y_{s'})(q^{2m} x_{r} \mu_r-x_s \mu_s)}{E_2(r,s,s')   (y_{s'}-x_r \mu_r\mu_{s'} q^{2(m+\ell)})}.\een
This statement can be proved by induction in $m$ and $\ell$ as above, with the choice 
\ben E_2(r,s,s')= \frac{\mu_{s'} q^{{2}}(z_r^2-z_{s'}^2)( x_{r} \mu_r-x_s \mu_s)}{ (y_{s'}-x_r \mu_r\mu_{s'})}.\een

\end{appendix}
\bibliographystyle{alpha}

\begin{thebibliography}{DJMM91b}

\bibitem[AV22a]{AV22b}
Andrea Appel and Bart Vlaar.
\newblock {Trigonometric K-matrices for finite-dimensional representations of quantum affine algebras}.
\newblock {\em J. Eur. Math. Soc}, pub online first, 2025.

\bibitem[AV22b]{AV22a}
Andrea Appel and Bart Vlaar.
\newblock {Universal K-matrices for quantum Kac-Moody algebras}.
\newblock {\em Representation Theory of the American Mathematical Society}, 26(26):764–824, 2022.

\bibitem[AV25]{AV24}
Andrea Appel and Bart Vlaar.
\newblock {Tensor K-matrices for quantum symmetric pairs}.
\newblock {\em Commun. Math. Phys.}, 406(5):100, 2025.

\bibitem[Bax89]{Baxter:1989vv}
R.~J. Baxter.
\newblock {Superintegrable chiral Potts model: Thermodynamic properties, an inverse model and a simple associated Hamiltonian}.
\newblock {\em J. Statist. Phys.}, 57:1--39, 1989.

\bibitem[BGK{\etalchar{+}}14]{BGKNR14}
Herman Boos, Frank G{\"o}hmann, Andreas Kl{\"u}mper, Khazret~S Nirov, and Alexander~V Razumov.
\newblock {Universal R-matrix and functional relations}.
\newblock {\em Reviews in Mathematical Physics}, 26(06):1430005, 2014.

\bibitem[BJM{\etalchar{+}}09]{BJMST09}
H.~Boos, M.~Jimbo, T.~Miwa, F.~Smirnov, and Y.~Takeyama.
\newblock Hidden {G}rassmann structure in the {XXZ} model {II}: creation operators.
\newblock {\em Commun. Math. Phys.}, 286(3):875--932, 2009.

\bibitem[BKMS90]{BKMS90}
V.~V. Bazhanov, R.~M. Kashaev, V.V. Mangazeev, and Yu.~G. Stroganov.
\newblock {$(Z_N\times)^{n-1}$ generalization of the chiral Potts model}.
\newblock {\em Comm. in Math. Physics}, 138:393--408, 1990.

\bibitem[B{\L}MS10]{BLMS10}
V.V. Bazhanov, T.~{\L}ukowski, C.~Meneghelli, and M.~Staudacher.
\newblock A shortcut to the {Q}-operator.
\newblock {\em Journal of Statistical Mechanics: Theory and Experiment}, 2010(11):P11002, 2010.

\bibitem[BLZ96]{BLZ96}
V.V. Bazhanov, S.L. Lukyanov, and A.B. Zamolodchikov.
\newblock Integrable structure of conformal field theory, quantum {K}d{V} theory and thermodynamic {B}ethe ansatz.
\newblock {\em Commun. in Mathematical Physics}, 177(2):381--398, 1996.

\bibitem[BLZ97]{BLZ97}
V.V. Bazhanov, S.L. Lukyanov, and A.B. Zamolodchikov.
\newblock Integrable {S}tructure of {C}onformal {F}ield {T}heory {II}. {Q}-operator and {DDV} equation.
\newblock {\em Commun. Math. Phys.}, 190:247--278, 1997.

\bibitem[BPAY88]{BPY88}
R.J. Baxter, J.H.H. Perk, and H.~Au-Yang.
\newblock {New solutions of the star-triangle relations for the chiral Potts model}.
\newblock {\em Physics Letters A}, 128(3):138--142, 1988.

\bibitem[BS90]{BS90}
VV~Bazhanov and Yu~G Stroganov.
\newblock Chiral potts model as a descendant of the six-vertex model.
\newblock {\em Journal of Statistical Physics}, 59:799--817, 1990.

\bibitem[BT18]{BTs18}
P.~Baseilhac and Z.~Tsuboi.
\newblock Asymptotic representations of augmented q-{O}nsager algebra and boundary {K}-operators related to {B}axter {Q}-operators.
\newblock {\em Nucl. Phys. B}, 929:397--437, 2018.

\bibitem[CVW24]{CVW24}
Alec Cooper, Bart Vlaar, and Robert Weston.
\newblock {A Q-operator for open spin chains II: boundary factorization}.
\newblock {\em Commun. in Mathematical Physics}, 405(5):110, 2024.

\bibitem[Dav94]{Davies94}
Philip~J. Davis.
\newblock {\em Circulant matrices}.
\newblock Chelsea, New York, 2nd ed. edition, 1994.

\bibitem[DCK92]{DCK92}
C. De Concini and V.G. Kacs.
\newblock  Representations of quantum groups at roots of unity.
\newblock {\em Progress in Mathematics}, 92:471-506, 1992.

\bibitem[DFM01]{DFM01}
T. Deguchi, K. Fabricius and B.M. McCoy. 
\newblock The $sl_2 $ Loop Algebra Symmetry of the Six-Vertex Model at Roots of Unity.
\newblock {\em J. Stat. Phys}, 102:701-736, 2001.

\bibitem[Der07]{De07}
S.E. Derkachov.
\newblock Factorization of the {R}-matrix. {I}.
\newblock {\em J. of Mathematical Sciences}, 143(1):2773--2790, 2007.

\bibitem[Der08]{De05}
S.E. Derkachov.
\newblock {Factorization of R-matrix and Baxter's Q-operator}.
\newblock {\em J. Math. Sci.}, 151:2880--2893, 2008.

\bibitem[DJMM91a]{DJMM91a}
Etsuro Date, Michio Jimbo, Kei Miki, and Tetsuji Miwa.
\newblock {Cyclic Representations of $U_q (\mathfrak{sl}(n+ 1, C))$ at $q^N= 1$}.
\newblock {\em Publications of the Research Institute for Mathematical Sciences}, 27(2):347--366, 1991.

\bibitem[DJMM91b]{DJMM91b}
Etsuro Date, Michio Jimbo, Kei Miki, and Tetsuji Miwa.
\newblock {New R Matrices Associated with Cyclic Representations of $U_q(A^2 (2))$}.
\newblock {\em Publications of the Research Institute for Mathematical Sciences}, 27(4):639--655, 1991.

\bibitem[DJMM91c]{DJMM91c}
Etsurō Date, Michio Jimbo, Kei Miki, and Tetsuji Miwa.
\newblock {Generalized chiral Potts models and minimal cyclic representations of $U_q(\widehat{\mathfrak{gl}}(n,{\bf C}))$}.
\newblock {\em Commun. in Mathematical Physics}, 137(1):133 -- 147, 1991.

\bibitem[DKK06]{DKK06}
S.E. Derkachov, D.~Karakhanyan, and R.~Kirschner.
\newblock {Baxter Q-operators of the XXZ chain and R-matrix factorization}.
\newblock {\em Nucl. Phys. B}, 738(3):368--390, 2006.

\bibitem[FH15]{FH15}
E.~Frenkel and D.~Hernandez.
\newblock Baxter's {R}elations and {S}pectra of {Q}uantum {I}ntegrable {M}odels.
\newblock {\em Duke Math. J.}, 164(12):2407--2460, 2015.

\bibitem[FH24]{FH24}
Edward Frenkel and David Hernandez.
\newblock {Extended Baxter relations and QQ-systems for quantum affine algebras}, 2024.
\newblock arXiv preprint arXiv:2312.13256.

\bibitem[FJM24]{FJM24}
B.~Feigin, M.~Jimbo, and E.~Mukhin.
\newblock {Remarks on $q$-difference opers arising from quantum toroidal algebras}, 2024.
\newblock arXiv preprint arXiv:2406.07265.

\bibitem[FS15]{FSz15}
R.~Frassek and I.~Sz\'{e}cs\'{e}nyi.
\newblock Q-operators for the open {H}eisenberg spin chain.
\newblock {\em Nucl. Phys. B}, 901:229--248, 2015.

\bibitem[HJ12]{HJ12}
D.~Hernandez and M.~Jimbo.
\newblock Asymptotic representations and Drinfeld rational fractions.
\newblock {\em Compositio Mathematica}, 148(5):1593--1623, 2012.

\bibitem[H20]{H20}
D.~Hernandez.
\newblock Representations of shifted quantum affine algebras.
\newblock {\em International Mathematics Research Notices}, 2023(11):11035-11126, 2023.



\bibitem[IU92]{IU92}
I.T. Ivanov and D.B. Uglov.
\newblock R-matrices for the semicyclic representations of $u_q(sl(2))$.
\newblock {\em Physics Letters A}, 167(5):459--464, 1992.

\bibitem[JMS21]{JMS21}
Michio Jimbo, Tetsuji Miwa, and Fedor Smirnov.
\newblock {\em Local operators in integrable models. {I}}, volume 256 of {\em Mathematical Surveys and Monographs}.
\newblock American Mathematical Society, Providence, RI, 2021.

\bibitem[Kor03a]{Korff_2003a}
Christian Korff.
\newblock Auxiliary matrices for the six-vertex model at $q^n = 1$ and a geometric interpretation of its symmetries.
\newblock {\em Journal of Physics A: Mathematical and General}, 36(19):5229, 2003.

\bibitem[Kor03b]{Korff_2003b}
Christian Korff.
\newblock {Auxiliary matrices for the six-vertex model at $q^N = 1$: II. Bethe roots, complete strings and the Drinfeld polynomial}.
\newblock {\em Journal of Physics A: Mathematical and General}, 37(2):385, 2003.

\bibitem[Kor04a]{Korff_2004a}
Christian Korff.
\newblock {The twisted XXZ chain at roots of unity revisited}.
\newblock {\em Journal of Physics A: Mathematical and General}, 37(5):1681, 2004.

\bibitem[Kor04b]{Korff_2004b}
Christian Korff.
\newblock {Auxiliary matrices for the six-vertex model and the algebraic Bethe ansatz}.
\newblock {\em Journal of Physics A: Mathematical and General}, 37:7227-7254, 2004.

\bibitem[KRS81]{KR81}
Petr~P Kulish, N~Yu Reshetikhin, and Evgeny~K Sklyanin.
\newblock Yang-baxter equation and representation theory: I.
\newblock {\em Letters in Mathematical Physics}, 5:393--403, 1981.

\bibitem[KT14]{KT14}
S.~Khoroshkin and Z.~Tsuboi.
\newblock The universal {R}-matrix and factorization of the {L}-operators related to the {B}axter {Q}-operators.
\newblock {\em J. Phys. A: Math. Theor.}, 47(19):192003, 2014.

\bibitem[MLP21]{MLP21}
Yuan Miao, Jules Lamers, and Vincent Pasquier.
\newblock {On the Q operator and the spectrum of the XXZ model at root of unity}.
\newblock {\em SciPost Phys.}, 11:067, 2021.

\bibitem[MNP18]{Maillet2018}
Jean~Michel Maillet, Giuliano Niccoli, and Baptiste Pezelier.
\newblock Transfer matrix spectrum for cyclic representations of the 6-vertex reflection algebra ii.
\newblock {\em SciPost Physics}, 5(3), 2018.

\bibitem[Nic10]{Niccoli2010}
G.~Niccoli.
\newblock {Reconstruction of Baxter -operator from Sklyanin SOV for cyclic representations of integrable quantum models}.
\newblock {\em Nuclear Physics B}, 835(3):263–283, 2010.

\bibitem[Roa06a]{Roan2007}
Shi-shyr Roan.
\newblock {\em Bethe Ansatz and Symmetry in Superintegrable Chiral Potts Model and Root-of-unity Six-vertex Model, Pub in book Differential Geometry and Physics}, pages 399--409.
\newblock World Scientific, 2006.

\bibitem[Roa06b]{Roan_2006}
Shi-shyr Roan.
\newblock {The Q-operator for root-of-unity symmetry in the six-vertex model}.
\newblock {\em Journal of Physics A: Mathematical and General}, 39(40):12303–12325, 2006.

\bibitem[Roa07]{Roan_2007}
Shi-shyr Roan.
\newblock {The transfer matrix of a superintegrable chiral Potts model as the Q-operator of root-of-unity XXZ chain with cyclic representation of $U_{\mathsf {q} }(sl_2)$}.
\newblock {\em Journal of Statistical Mechanics: Theory and Experiment}, 2007(09):P09021–P09021, 2007.

\bibitem[Roa09]{Roan_2009}
Shi-shyr Roan.
\newblock {On the $\tau(2)$-model in the chiral Potts model and cyclic representation of the quantum group $U_q(sl_2)$}.
\newblock {\em Journal of Physics A: Mathematical and Theoretical}, 42(7):072003, 2009.

\bibitem[Tsu21]{Tsuboi2021}
Zengo Tsuboi.
\newblock {Universal Baxter TQ-relations for open boundary quantum integrable systems}.
\newblock {\em Nuclear Physics B}, 963:115286, February 2021.

\bibitem[VW20]{VW20}
Bart Vlaar and Robert Weston.
\newblock {A Q-operator for open spin chains I. Baxter’s TQ relation}.
\newblock {\em Journal of Physics A: Mathematical and Theoretical}, 53(24):245205, 2020.

\end{thebibliography}
\newcommand{\etalchar}[1]{$^{#1}$}

\end{document}